\makeatletter
\newcommand{\dontusepackage}[2][]{%
  \@namedef{ver@#2.sty}{9999/12/31}%
  \@namedef{opt@#2.sty}{#1}}
\makeatother
\dontusepackage{subfigure}
\documentclass{article}
\usepackage{amssymb,amsmath,amsbsy,bm}
\usepackage{fixltx2e} 
\IfFileExists{upquote.sty}{\usepackage{upquote}}{}
\usepackage[utf8]{inputenc}
\IfFileExists{microtype.sty}{\usepackage{microtype}}{}
\usepackage{natbib}
\usepackage{listings}
\usepackage{graphicx}
\makeatletter
\def\ScaleIfNeeded{%
  \ifdim\Gin@nat@width>\linewidth
    \linewidth
  \else
    \Gin@nat@width
  \fi
}
\makeatother
\let\Oldincludegraphics\includegraphics
{
 \catcode`\@=11\relax
 \gdef\includegraphics{\@ifnextchar[{\Oldincludegraphics}{\Oldincludegraphics[width=\ScaleIfNeeded]}}
}
\usepackage{caption}
\usepackage{float}
\usepackage{subfig}
\usepackage{algorithm} 
\let\scholmdAlgorithm\algorithm
\let\endscholmdAlgorithm\endalgorithm
\let\algorithm\relax \let\endalgorithm\relax

{
 \catcode`\*=11\relax
 \global\let\scholmdAlgorithm*\algorithm*
 \global\let\endscholmdAlgorithm*\endalgorithm*
 \global\let\algorithm*\relax 
 \global\let\endalgorithm*\relax
}

\usepackage[unicode=true]{hyperref}
\hypersetup{breaklinks=true,
            bookmarks=true,
            pdfauthor={},
            pdftitle={A dual formulation of wavefield reconstruction inversion for large-scale seismic inversion},
            colorlinks=true,
            citecolor=black,
            urlcolor=blue,
            linkcolor=black,
            pdfborder={0 0 0}}
\usepackage{mathtools}
\newcommand{\bmm}{\mathbf{m}}

\newcommand{\bx}{\mathbf{x}}
\newcommand{\bu}{\mathbf{u}}
\newcommand{\bw}{\mathbf{w}}
\newcommand{\bub}{\bar{\bu}}
\newcommand{\bd}{\mathbf{d}}
\newcommand{\bq}{\mathbf{q}}
\newcommand{\bfb}{\mathbf{f}}
\newcommand{\by}{\mathbf{y}}
\newcommand{\br}{\mathbf{r}(\bmm)}
\newcommand{\bv}{\mathbf{v}}
\newcommand{\F}{F(\bmm)}
\newcommand{\A}{A(\bmm)}
\newcommand{\R}{R}
\newcommand{\Real}{\mathbb{R}}
\newcommand{\Compl}{\mathbb{C}}
\newcommand{\J}{\mathcal{J}}
\newcommand{\Jpen}{\J_{\mathrm{pen}}}
\newcommand{\JFWI}{\J_{\mathrm{FWI}}}
\newcommand{\JWRI}{\J_{\mathrm{WRI}}}

\renewcommand{\L}{\mathcal{L}}
\newcommand{\Lwav}{\L_{\mathrm{weq}}}
\newcommand{\Ldat}{\L_{\mathrm{dat}}}
\newcommand{\LWRId}{\L_{\mathrm{WRI}^{\ast}}}

\newcommand{\LWRIdl}{\L_{\mathrm{WRI}^{\ast},\lambda}}
\newcommand{\LWRIdg}{\widetilde{\L}_{\mathrm{WRI}^{\ast}}}
\newcommand{\norm}[1]{\left\lVert#1\right\rVert}
\renewcommand{\ast}{^*}
\newcommand{\Nr}{N_{\mathrm{r}}}
\newcommand{\Ns}{N_{\mathrm{s}}}
\newcommand{\Nf}{N_{\mathrm{f}}}
\newcommand{\Nt}{N_{\mathrm{t}}}

\newcommand{\<}{\langle}
\renewcommand{\>}{\rangle}
\newcommand{\defeq}{:=}

\title{A dual formulation of wavefield reconstruction inversion for large-scale
seismic inversion}
\author{Gabrio Rizzuti\textsuperscript{1} \and Mathias Louboutin\textsuperscript{1} \and
Rongrong Wang\textsuperscript{2} \and Felix J.
Herrmann\textsuperscript{1}\\\textsuperscript{1} Georgia Institute of
Technology,\\\textsuperscript{2} Michigan State University}
\date{}

\begin{document}
\maketitle

\section{Abstract}\label{abstract}

Many of the seismic inversion techniques currently proposed that focus
on robustness with respect to the background model choice are not apt to
large-scale 3D applications, and the methods that are computationally
feasible for industrial problems, such as full waveform inversion, are
notoriously bogged down by local minima and require adequate starting
models. We propose a novel solution that is both scalable and less
sensitive to starting model or modeling inaccuracies (e.g.~stemming from
inaccurate physical parameters that are kept fixed during inversion)
when compared to full waveform inversion. It is based on a dual
reformulation of the classical wavefield reconstruction inversion, whose
empirical robustness with respect to local minima is well documented in
the literature. Alas, the classical version is not suited to 3D, as it
leverages expensive frequency-domain solvers for the wave equation. Our
proposal, instead, allows the deployment of state-of-the-art time-domain
finite-difference methods, and is computationally mature for industrial
scale problems.

\section{Introduction}\label{introduction}

Field-data applications of wave-equation based seismic inversion are
limited by the computational complexity of large-scale 3D simulations,
and by the tendency of traditional inversion techniques \citep[such as
full waveform inversion, FWI,][]{tarantola1982generalized} to converge
to unrealistic models. While considerable advances have been made to
speed up wave equation solvers and avoid local minima, most of the
currently proposed inversion methods do not take advantage of both these
developments. While FWI is applied
routinely in industry, techniques that are robust to local minima are
typically limited to small problems. In
this paper, we introduce a novel method that is fit for 3D large-scale
deployment and relatively robust against spurious minima.

We adopt model extension methods that aim at counteracting local minimum
stagnation, which are based on a reformulation of the original inverse
problem into a higher dimensional space of unknowns \citep{symesMVAFWI}.
While beneficial for avoiding local minima, the inherent added
complexity of these methods typically curtails large-scale applications.
One notable example, and the basis for this work, is wavefield
reconstruction inversion \citep[WRI,][]{van2013mitigating}, which has
demonstrated robustness towards local minima but is not well suited for
3D problems. Indeed, WRI involves the solution of an extended wave
equation that is conventionally solved in the frequency domain with
direct or iterative methods, although not efficiently for large grid
sizes. Many other methods related to WRI share similar limitations
\citep{huang2017full, huang2018volume, aghamiry2019improving}.

We propose a principled Lagrangian reformulation of classical WRI,
referred to as WRI$\ast$ throughout the text, which naturally lends
itself to time-domain solvers, hence large-scale applications. The
reformulation is based on minimizing the wave equation error and
interpreting the data misfit as a constraint.
The resulting method has comparable
computational costs to FWI, since it is based on a objective functional
whose gradient computation involves twice as many physical simulations
required by FWI. Moreover, it is consistently more robust than FWI with
respect to starting models and inexact modeling parameters not subject
to inversion. Our approach represents a
computationally efficient, robust, and ultimately feasible proposal for
3D seismic inversion.

Our proposal leverages current state-of-the-art in time-domain
finite-difference solvers that ensures scalability on the hardware
available to the oil-and-gas industry, or even to a wider audience
thanks to cloud computing \citep{witte2019TPDedas}. We employ the
domain-specific language Devito for automatic generation of highly
optimized finite-difference C code
\citep{luporini2018aap, luporini2018architecture, louboutin2018devito}.
Here, we focus on acoustic modeling and pseudo-acoustic modeling with
tilted transverse isotropy \citep[TTI,][]{alkhalifah2000}, established
as a reasonable compromise between realistic physics and computational
feasibility.

The computational and theoretical advantages of WRI$\ast$ will be
demonstrated with extensive 2D experimentation. Despite the central
claim of the paper being centered on 3D feasibility, realistic 3D
problems are still beyond the economic resources realistically available
to academic organizations, and will not be included in this work.
Other fields relying on heavy
computations, such as machine learning on large models, are plagued by similar problems.
Therefore, we will present some preliminary results for a small 3D
example. We stress that this is not due to
a fundamental limitation of the method, and any institution that can
afford the computations needed for FWI can also fruitfully employ
WRI$\ast$.

In the rest of this paper, we discuss the computational and theoretical
pitfalls of classical seismic inversion, with a general overview of the
methods that deal with those issues. We review Lagrangian and reduced
formulations for seismic inversion, followed by a more in-depth
discussion on WRI. The main section is devoted to the reformulation of
WRI object of this work: WRI$\ast$. Lastly, we present numerical
experiments aiming at the comparison with FWI and conventional WRI.

\section{Seismic inversion: computational and theoretical
complexity}\label{seismic-inversion-computational-and-theoretical-complexity}

Modern seismic inversion, such as FWI, is cast as an optimization
problem endowed with partial differential equation constraints
\citep{virieux2009overview}. Given a suitable discretization of the
physical quantities of interest, and the algebraic operators that govern
their mutual relationship, the problem can be succinctly expressed as:
\begin{equation}
\min_{\bmm,\bu}\dfrac{1}{2}\norm{\bd-\R\bu}^2\quad\mathrm{subject\ to}\quad\A\bu=\bq.
\label{eq:invprobgen}
\end{equation}
 Data $\bd\in\Real^{\Nr\times\Nt\times\Ns}$ are recorded in time (where
$Nt$ denotes the number of time samples) at $\Nr$ receiver locations for
different $\Ns$ source positions. Alternatively, after a temporal
Fourier transform, $\bd\in\Compl^{\Nr\times\Nf\times\Ns}$ for $\Nf$
frequencies. The variable $\bu\in\Real^{N\times\Nt\times\Ns}$ (or
$\bu\in\Compl^{N\times\Nf\times\Ns}$, in frequency domain) represents a
wavefield---e.g.~the solution of the wave equation $\A\bu=\bq$---
discretized on an $N$-dimensional space. The right-hand side $\bq$ of
the wave equation is typically a point source. In
equation~\eqref{eq:invprobgen}, the data are compared to the values of
$\bu$ interpolated at the receiver locations via the operator $R$, and
the misfit is calculated via the least-squares norm
$\norm{\cdot}\defeq\norm{\cdot}_2$. The wave operator $\A$ is
parameterized by $\bmm$, which encapsulates the model unknowns of the
problem. Throughout this paper, $\bmm$ represents squared slowness.
Other physical parameters are included in the wave operator but kept
fixed and will not be estimated.

In 3D, the grid size grows as $N=\mathcal{O}(n^3)$ (in big $\mathcal{O}$
notation), with $n\approx1000$ grid points in each direction for a
typical industrial-scale problem. With dense source/receiver acquisition
coverage, $\Ns=\mathcal{O}(n^2)$ and $\Nr=\mathcal{O}(n^2)$.
Furthermore, when time-domain methods with explicit time-marching
schemes are applied, $\Nt=\mathcal{O}(n)$ (realistically,
$\Nt\approx10\,n$), due to stability conditions. In the frequency
domain, we also often keep $\Nf=\mathcal{O}(n)$, albeit with much more
favorable complexity constants than time domain
\citep[as suggested
in][]{sirgueFreqStrategy}. In this regime,
the computational requirements are extremely challenging. For instance,
wavefield storage grows as $\mathcal{O}(n^6)$, and cannot be attained
for each source/time sample at once. Time complexity is largely
dominated by the computation of the solutions of the wave equation.
Currently, explicit time-domain methods are favored, as in the frequency
domain the wave equation is represented by a large, sparse, non-definite
linear system, and neither direct nor iterative methods scale as
efficiently to 3D. Frequency domain becomes a palatable option only when
a large high-performance computing system can store the matrix
factorization \citep{wang20113d}, or abundant compute cores are
available \citep{knibbe2016reduction}.

Despite these computational challenges, FWI is now viable for 3D
problems of practical size, especially when the simulated physics is
restricted to acoustics or tilted transversely isotropic (TTI)
pseudo-acoustics \citep{alkhalifah2000, grechka2004, duveneck2008}. The
FWI objective functional is, however, highly multimodal and requires
considerable problem-specific intervention to be consistently successful
when gradient-based optimization is employed. In practice, FWI requires
a kinematically correct starting guess to avoid cycle skipping. Local
minima have been the subject of a great deal of research in the last 10
years, and many proposals tackling the problem have been advanced. The
main lines of investigation involved:

\begin{itemize}
\itemsep1pt\parskip0pt\parsep0pt
\item
  regularization techniques that penalize unrealistic models, e.g.~via
  total-variation minimization
  \citep{akcelik2002parallel, anagaw2012edge, esser2016tvr} or
  projection onto constraint sets \citep{peters2016cvp, peters2018pmf};
\item
  analysis of data misfit in a transformed domain
  \citep{shinFWIlapl, shinFWIlaplfour, bozdagFWIhilb} and/or application
  of preprocessing hierarchical strategies, where different levels of
  data preconditioning are setup and the associated problems solved
  sequentially, e.g.~low- to high-frequency data filtering inversion
  \citep{bunksFWI}, or shallow- to deep-model inversion
  \citep{shippLayer};
\item
  model extension, where the original problem is generalized to a higher
  dimensional space better suited for gradient-based optimization
  \citep{symesMVAFWI, van1997contrast, biondiTFWI2, van2013mitigating, wang2016full, wang2017denoising, aghamiry2019improving, warnerAWI1, warnerAWI2};
\item
  special objective functionals more amenable to local-search
  optimization, e.g.~based on traveltime-based misfit
  \citep{luo91, van2010correlation}, or optimal transport via the
  Wasserstein distance
  \citep{metivierFWIopttransp, yang2016application, Sun2019}.
\end{itemize}

In practice, most of the inversion techniques typically benefit from a
combination of these strategies.

Recently, many model extension methods have been proven successful in
their empirical robustness to local minima, but are hampered by the
computational costs associated to operating in higher dimension. A
notable exception is adaptive waveform inversion
\citep[AWI,][]{warnerAWI1, warnerAWI2}, which consists of matching
synthetic time traces to data by means of convolutional Wiener filters
with arbitrary length. Since the computational overhead with respect to
FWI is affordable, AWI can be applied to large problems. In this paper,
however, we are not interested in establishing the relative merits of
AWI with respect to other methods, and we will focus on developing a
competing method based on WRI. While
theoretically not immune to cycle-skipping, as demonstrated in the
recent work \citet{symes2020waveform} under simplifying 1-D assumptions,
WRI is empirically more robust than FWI to local minima \citep[as
evidenced by a wealth of publications, starting from its original
inception in][to this very paper]{van2013mitigating}. Moreover, as
suggested in \citet{symes2020waveform}, a relatively straightforward
modification of the WRI objective (based on a specific weighting of the
equation error misfit) is already proven to be effective, albeit under
the same simplifying conditions. Note that these modifications of WRI
are readily implemented in this treatise.
Despite these advantages, WRI in its basic formulation
\citep{van2013mitigating} is not amenable to 3D. Indeed, it involves a
relaxed version of the wave equation that does not allow for a
straightforward implementation of time-marching schemes and is
practically limited to 2D problems \citep[see, however,][ for some
advances using frequency-domain 3D solvers]{peters2019SEGans}.
Alternative characterizations of the original WRI formulation, as
offered in \citet{huang2017full}, \citet{huang2018volume}, or
\citet{aghamiry2019improving}, suffer from similar computational issues
and do not scale efficiently to 3D. Note that the approaches of \citet{huang2017full},
\citet{huang2018volume} can be cast in the time domain, but still
require expensive iterative solvers for the state variable, incurring
the same computational bottleneck of WRI, or even FWI when formulated in
the frequency domain. The main goal of
this paper is to provide a reformulation of WRI that circumvents this
limitation.

\section{Reduced, full-space, and penalty
formulations}\label{reduced-full-space-and-penalty-formulations}

In this section, we lay out the groundwork for WRI and our proposal
WRI$\ast$ by illustrating different ways to deal with the problem
formulated in equation~\eqref{eq:invprobgen}. Classical FWI consists of
imposing the constraint $\bu=\A^{-1}\bq$ explicitly (from which the
moniker of reduced formulation stems). Equation~\eqref{eq:invprobgen}
can be equivalently rewritten as:
\begin{equation}
\begin{split}
& \min_{\bmm}\JFWI(\bmm),\\
& \JFWI(\bmm)=\dfrac{1}{2}\norm{\bd-\R\A^{-1}\bq}^2.
\end{split}
\label{eq:FWI}
\end{equation}
 Regularization can be enforced via constraints or extra penalty terms.
There are several advantages to this formulation that makes large-scale
deployment possible with the computational capabilities that are
currently available. The objective functional $\JFWI$ is defined by a
summation of the data misfit terms over individual sources. Therefore,
in the evaluation of $\JFWI$ and its gradient calculation, the solution
wavefields of forward and adjoint problems pertaining to a source can be
discarded after use, thus limiting the memory overhead. In time domain,
checkpointing techniques also reduce
memory complexity
\citep{symes2007reverse}. Alternatively, one could combine modeling in
the time domain and the on-the-fly Fourier transform to obtain an
estimate of the gradient by cross-correlating the forward and
backpropagated wavefields in the Fourier domain
\citep{sirgue20083d, witte2018cls}. Although the reduced formulation
leads to a scalable approach by making use of relatively frugal
time-domain modeling, it is particularly prone to local minima, and
necessitates a kinematically adequate starting guess for $\bmm$, ideally
matching the data traveltimes within half a period \citep{Beydoun1988}.

Another classical setup, sometimes referred to as the full-space method,
is by means of the Lagrangian function associated to
equation~\eqref{eq:invprobgen} and the related saddle point problem
\citep{haber2000optimization}:
\begin{equation}
\begin{split}
& \max_{\bv}\min_{\bmm,\bu}\Lwav(\bmm,\bu,\bv),\\
& \Lwav(\bmm,\bu,\bv)=\dfrac{1}{2}\norm{\bd-\R\bu}^2+\<\bv,\bq-\A\bu\>,
\end{split}
\label{eq:lagrweq}
\end{equation}
 where $\<\cdot,\cdot\>$ is the least-squares inner product. The
multipliers $\bv$ belong to a linear space with the same dimension as
the wavefield $\bu$ space. When memory usage is not of concern, we can
carry out joint optimization over $\bmm$, $\bu$ and $\bv$. Note that the
evaluation and gradient computation of the Lagrangian do not require the
solution of the wave equation. Moreover, the Hessian of the Lagrangian
is sparse and can be evaluated cheaply, making second-order methods
affordable. However, in spite of these advantages, storing $\bu$ and
$\bv$ in memory is not viable.

Penalty formulations are situated in between the full space method and
FWI, and seek to enforce the wave equation weakly as a regularization
term. By fixing the multiplier $\bv=\lambda^2(\bq-\A\bu)$ in
equation~\eqref{eq:lagrweq}, we obtain the problem:
\begin{equation}
\begin{split}
& \min_{\bmm,\bu}\Jpen(\bmm,\bu),\\
& \Jpen(\bmm,\bu)=\dfrac{1}{2}\norm{\bd-\R\bu}^2+\dfrac{\phantom{^2}\lambda^2}{2}\norm{\bq-\A\bu}^2.
\end{split}
\label{eq:pen}
\end{equation}
 Here, $\lambda$ is a trade-off parameter between data misfit and wave
equation error. Note that, with $\lambda\rightarrow\infty$, the method
becomes equivalent to FWI. Similarly to the full space method, we cannot
afford the storage of all the unknowns $\bu$, and need to resort to the
WRI approach described in the next section.

\section{Wavefield reconstruction
inversion}\label{wavefield-reconstruction-inversion}

Within the penalty method framework in equation~\eqref{eq:pen},
\citet{van2013mitigating} apply the variable projection scheme by
defining $\bar{\bu}(\bmm)\defeq\arg\min_{\bu}\Jpen(\bmm,\bu)$
\citep{golub2003separable}. This procedure yields the WRI method, whose
reduced objective, function of only model variables $\bmm$, is:
\begin{equation}
\begin{split}
& \min_{\bmm}\JWRI(\bmm),\\
& \JWRI(\bmm)=\dfrac{1}{2}\norm{\bd-\R\bub(\bmm)}^2+\dfrac{\phantom{^2}\lambda^2}{2}\norm{\bq-\A\bub(\bmm)}^2,
\end{split}
\label{eq:WRI}
\end{equation}
 where $\bub$ is the solution of the augmented wave equation:
\begin{equation}
\begin{split}
\begin{pmatrix}
\R\\
\lambda\A\\
\end{pmatrix}\bub(\bmm)=
\begin{pmatrix}
\bd\\
\lambda\bq\\
\end{pmatrix},
\end{split}
\label{eq:WRIweq}
\end{equation}
 to be solved in least-squares sense. Equivalently,
\begin{equation}
(\R^{\ast}\R+\lambda^2\A^{\ast}\A)\bub(\bmm)=\R^{\ast}\bd+\lambda^2\A^{\ast}\bq.
\label{eq:WRIweqAlt}
\end{equation}
 Note that, by virtue of variable projection,
$\nabla_{\bu}\Jpen(\bmm,\bub(\bmm))=\mathbf{0}$, and
\begin{equation}
\nabla_{\bmm}\JWRI(\bmm)=\nabla_{\bmm}\Jpen(\bmm,\bub(\bmm)).
\label{eq:gradWRI}
\end{equation}
 As in the reduced approach, WRI avoids simultaneous wavefield storage
for each source (by computing and discarding). Empirical indications
(including this paper) show that WRI is more robust than FWI with
respect to local minima. Furthermore, WRI delivers consistently superior
salt imaging \citep[especially when combined with
regularization,][]{esser2016tvr, peters2016cvp, da2017wavefield}. Alas,
a major drawback of WRI is that the augmented wave
equation~\eqref{eq:WRIweq} cannot be trivially solved in neither
frequency nor time domain when the problem is sizable.

Different reformulations of WRI have
been proposed over the years that seek to overcome some of the
computational limitations just described. In the following, we discuss
some technical aspects related to the extended-source reformulation of
WRI, object of the \citet{wang2016full} and \citet{huang2018volume}
studies.

\citet{wang2016full} reformulate WRI
in terms of extended source variables rather than wavefields $\bu$, via
the change of variables $\bq_{\mathrm{ext}}\defeq\A\bu$. Rewriting
equation~\eqref{eq:pen} in terms of $\bq_{\mathrm{ext}}$, we
obtain:
\begin{equation}
\Jpen^q(\bmm,\bq_{\mathrm{ext}})=\dfrac{1}{2}\norm{\bd-\F\bq_{\mathrm{ext}}}^2+\dfrac{\phantom{^2}\lambda^2}{2}\norm{\bq-\bq_{\mathrm{ext}}}^2.
\label{eq:penvol}
\end{equation}
where the forward operator is defined
by
\begin{equation}
\F=\R\A^{-1}.
\label{eq:fwop}
\end{equation}
 Following the WRI approach, the
extended source variables might be solved via variable projection, e.g.
$(\F^*\F+\lambda^2 I)\bar{\bq}_{\mathrm{ext}}(\bmm)=\F^*\bd+\lambda^2\bq$.
To avoid the need for solving this linear system, the following
approximation is introduced:
\begin{equation}
\bar{\bq}_{\mathrm{ext}}(\bmm)\approx\tilde{\bq}_{\mathrm{ext}}(\bmm)=\bq+\dfrac{1}{\lambda^2}\F^*\br,\quad\br=\bd-\F\bq.
\label{eq:wang}
\end{equation}
Here, $\br$ is the data misfit
residual for the model $\bmm$. The approximation in
equation~\eqref{eq:wang} can be regarded as the first iteration of a
Krylov subspace solver applied to the system $\F^*\F+\lambda^2 I$, and
its accuracy increases as $\lambda\to\infty$ (which reveals the weakness
of the approximation, since for $\lambda\to\infty$ WRI reverts to FWI).
The computations related to equation~\eqref{eq:wang} involve standard
wave equation solvers, even in time domain, and are fit to realistic
problems. A reduced objective is then
introduced:
\begin{equation}
\widetilde{\mathcal{J}}_{\mathrm{WRI}}^q(\bmm)=\dfrac{1}{2}\norm{\bd-\F\tilde{\bq}_{\mathrm{ext}}(\bmm)}^2+\dfrac{\phantom{^2}\lambda^2}{2}\norm{\bq-\tilde{\bq}_{\mathrm{ext}}(\bmm)}^2.
\label{eq:wang_obj}
\end{equation}
Since the extended source variable is
only approximated, here, the computation of the gradient of
$\widetilde{\mathcal{J}}_{\mathrm{WRI}}^q$ is not trivial, and
\citet{wang2016full} resort to the
estimation:
\begin{equation}
\nabla_{\bmm}\widetilde{\mathcal{J}}_{\mathrm{WRI}}^q(\bmm)\approx\nabla_{\bmm}\Jpen^q(\bmm,\tilde{\bq}_{\mathrm{ext}}(\bmm)).
\label{eq:wang_grad}
\end{equation}
Note that the dependence of
$\widetilde{\bq}_{\mathrm{ext}}(\bmm)$ with respect to $\bmm$ is not
properly taken into account. The overall approach can indeed tackle
large scale inversion, but is defective in two ways: the extended source
approximation in equation~\eqref{eq:wang} can be scaled badly and
gradient-based optimization can suffer from the inaccurate gradient
calculations involved in
equation~\eqref{eq:wang_grad}.

Volume source-based extended waveform
inversion, by \citet{huang2018volume}, closely follow the extended
source reformulation of \citet{wang2016full}, although via a slightly
different change of variable formula
$\bq_{\mathrm{ext}}\defeq\bq-\A\bu$. For simplicity, we will discuss the
work of \citet{huang2018volume} with the extended source definition
given by \citet{wang2016full}. Contrary to \citet{wang2016full}, here
the augmented unknowns are solved via variable projection, and,
analogously to WRI, a reduced objective can be defined only in terms of
model variables $\bmm$ (therefore without suffering from the gradient
approximation introduced in equation~\ref{eq:wang_grad}). One of the
central contributions of \citet{huang2018volume} is to refine the
regularization term in~\eqref{eq:penvol} with prior knowledge about the
source location $\bx_s$, e.g.~via the weighted
penalty
\begin{equation}
\Jpen^{q,\Sigma}(\bmm,\bq_{\mathrm{ext}})=\dfrac{1}{2}\norm{\bd-\F\bq_{\mathrm{ext}}}^2+\dfrac{\phantom{^2}\lambda^2}{2}\norm{\bq-\bq_{\mathrm{ext}}}_{\Sigma}^2,
\label{eq:penvol_src}
\end{equation}
defined by the weighted least-squares
norm
$\norm{\bq_{\mathrm{ext}}}_{\Sigma}\defeq\sqrt{\<\Sigma^{-1}\bq_{\mathrm{ext}},\bq_{\mathrm{ext}}\>}$.
The weighting matrix $\Sigma^{-1}$ is chosen to penalize the
contribution of the extended source located far from the physical source
position, by imposing a diagonal structure
$\Sigma^{-1}=\mathrm{diag}(\bw)$ corresponding to the spatial
function:
\begin{equation}
\bw(\bx)=|\bx-\bx_{\mathrm{s}}|.
\label{eq:annihilator}
\end{equation}
The relevance of the weighted
regularization on suppressing local minima has been recently discussed
in \citet{symes2020wavefield} and \citet{symes2020waveform}, where
conventional WRI with no weighting has been shown to exhibit the same
local minima of FWI for some 1-D examples.

The extended source variables are
functions of both space and time, and cannot be feasibly kept in memory
all at once. Therefore, \citet{huang2018volume} propose to restrict
those unknowns to space only. While volume source-based extended
waveform inversion is memory efficient, variable projection requires
iterative solutions of a linear system akin to
equation~\eqref{eq:WRIweq}, hence sharing the same computational
bottleneck as WRI.

\section{\texorpdfstring{A dual formulation for wavefield reconstruction
inversion:
WRI$\ast$}{A dual formulation for wavefield reconstruction inversion: WRI\textbackslash{}ast}}\label{a-dual-formulation-for-wavefield-reconstruction-inversion-wriast}

A way to overcome the computational hurdles associated to the augmented
wave equation is suggested by 
considering equation~\eqref{eq:WRIweqAlt}. Multiplying both sides of
this expression by the inverse of $\A^{\ast}$ and simple algebraic
manipulations yields
\begin{equation}
\A\bub(\bmm)=\bq+\bar{\bq}(\bmm),\quad\A^{\ast}\bar{\bq}(\bmm)=\R^{\ast}\bar{\by}(\bmm),
\label{eq:WRIweq2}
\end{equation}
 for some variable $\bar{\by}(\bmm)$ that belongs to the data space
(${\ast}$ denotes the adjoint operation). In
equation~\eqref{eq:WRIweq2}, $\bar{\by}(\bmm)$ is backpropagated to form
an additional source contribution $\bar{\bq}(\bmm)$. This extended
source has a time and spatial distribution and undergirds the idea of
extended-source inversion by \citet{wang2016full} and
\citet{huang2017full} (previously discussed). Moreover,
\begin{equation}
\bar{\by}(\bmm)=(\bd-\R\bub(\bmm))/\lambda^2.
\label{eq:WRIweq3}
\end{equation}
 This identity highlights that $\bar{\by}(\bmm)$ is proportional to the
data residual of the augmented wavefield defined in
equation~\eqref{eq:WRIweq}. It can be equivalently cast as the solution
of the linear system:
\begin{equation}
\left[\lambda^2 I+\F\F^{\ast}\right]\bar{\by}(\bmm)=\br.
\label{eq:yopt}
\end{equation}
 $\F$ is the forward operator defined in equation~\eqref{eq:fwop}, $I$
is the identity, and $\br=\bd-\F\bq$ is the data residual of the
physical solution of the wave equation. The evaluation of the linear
sytem in equation~\eqref{eq:yopt} involves solving the conventional wave
equation (and adjoint thereof). If one could cheaply estimate
$\bar{\by}(\bmm)$, the augmented solution $\bub(\bmm)$ would only
require two conventional wave equation solutions.

These insights follow from first principles by starting with a denoising
reformulation \citep{wang2017SEGdff} of equation~\eqref{eq:invprobgen},
\begin{equation}
\min_{\bmm,\bu}\dfrac{1}{2}\norm{\bq-\A\bu}^2\quad\mathrm{s.t.}\quad\norm{\bd-\R\bu}\leq\epsilon,
\label{eq:denoise}
\end{equation}
 for a known noise level $\epsilon$. Notice how objective and
constraints are here swapped with respect to the Lagrangian formulation
for the full-space method in equation~\eqref{eq:lagrweq}. The denoising
problem in equation~\eqref{eq:denoise} is somewhat equivalent to the
penalty formulation in equation~\eqref{eq:pen}, in the sense that for a
fixed $\bmm$ and a given $\epsilon$, there exists a weight $\lambda$
such that the solutions of~\eqref{eq:pen} and~\eqref{eq:denoise} are
identical \citep[and vice versa,][]{bjork96}. The denoising formulation,
however, has the advantage of a lower dimensional Lagrangian multiplier
than the penalty counterpart, as described in the following sections.

\subsection{Lagrangian formulation for data misfit
constraint}\label{lagrangian-formulation-for-data-misfit-constraint}

We proceed similarly to the full-space method, by building the
Lagrangian associated to the problem~\eqref{eq:denoise}. In order to do
so, we refer to the Fenchel duality theory
\citep{rockafellar1970convex}. The saddle-point problem associated to
equation~\eqref{eq:denoise} takes the form
\begin{equation}
\begin{split}
& \max_{\by}\min_{\bmm,\bu}\Ldat(\bmm,\bu,\by),\\
& \Ldat(\bmm,\bu,\by)=\dfrac{1}{2}\norm{\bq-\A\bu}^2+\<\by,\bd-\R\bu\>-\epsilon\norm{\by},
\end{split}
\label{eq:lagr}
\end{equation}
 where the slack variables $\by$ belong to a linear space with the same
dimension as data. We can eliminate the wavefield variables $\bu$ from
equation~\eqref{eq:lagr} by applying the variable projection
$\bar{\bu}(\bmm)=\arg\min_{\bu}\Ldat(\bmm,\bu,\by)$:
\begin{equation}
\A\bub(\bmm)=\bq+\bar{\bq}(\bmm),\quad\A^{\ast}\bar{\bq}(\bmm)=\R^{\ast}\by,
\label{eq:varproj}
\end{equation}
 in a similar fashion to what was described in
equation~\eqref{eq:WRIweq2}. The substitution
$\LWRId(\bmm,\by)\defeq\Ldat(\bmm,\bub(\bmm),\by)$ yields the WRI$\ast$
problem
\begin{equation}
\begin{split}
& \max_{\by}\min_{\bmm}\LWRId(\bmm,\by),\\
& \LWRId(\bmm,\by)=-\dfrac{1}{2}\norm{\F^{\ast}\by}^2+\<\by,\br\>-\epsilon\norm{\by},
\end{split}
\label{eq:lagr_red}
\end{equation}
 where the forward operator $\F=\R\A^{-1}$ was introduced in
equation~\eqref{eq:fwop}, and 
$\br=\bd-\F\bq$ is the data residual for
the model $\bmm$. In equation~\eqref{eq:lagr_red}, optimizing over $\by$
means maximizing the correlation of $\by$ with the data residual $\br$,
where additional regularization terms penalize the magnitudes of $\by$
and the extended source $\bar{\bq}(\bmm)=\F^{\ast}\by$. Under this lens,
the method bears some resemblance with techniques based on data
cross-correlation thoroughly investigated in the past starting from the
work of \citet{luo91}.

Regarding the gradient calculation, we
start by denoting with $J[\bmm,\bfb]=\partial_{\bmm}(\F\bfb)$ the
Jacobian of the mapping $\bmm\mapsto\F\bfb$, for a general source $\bfb$
(we make here a notational distinction between the derivative of a
vector-valued function, indicated by $\partial_{\bmm}$, and the
derivative of a real-valued function, indicated by $\nabla_{\bmm}$). The
action of the adjoint of $J[\bmm,\bfb]$ can be obtained by
differentiating both sides of the identity
$\<\by,\F\bfb\>=\<\F^{\ast}\by,\bfb\>$, which gives
$J[\bmm,\bfb]^{\ast}\by=\partial_{\bmm}(\F^{\ast}\by)^{\ast}\bfb$. This
promptly yields
$\nabla_{\bmm}\dfrac{1}{2}\norm{\F^{\ast}\by}^2=J[\bmm,\F^{\ast}\by]^{\ast}\by$
and
$\nabla_{\bmm}\<\by,\br\>=-J[\bmm,\bq]^{\ast}\by$.Therefore,
we obtain
\begin{gather}
\label{eq:lagr_red_gradm} \nabla_{\bmm}\LWRId=-J[\bmm,\bq+\F^{\ast}\by]^{\ast}\by,\\
\label{eq:lagr_red_grady} \nabla_{\by}\LWRId=\bd-\F(\bq+\F^{\ast}\by)-\epsilon\by/\norm{\by}.
\end{gather}
 The expression in equation~\eqref{eq:lagr_red_gradm} can be computed
similarly to a conventional FWI gradient: (\emph{i}) compute
$\bar{\bq}(\bmm)$ in equation~\eqref{eq:varproj} by backward
propagation, (\emph{ii}) compute $\bar{\bu}(\bmm)$ in
equation~\eqref{eq:varproj} by forward propagation, and (\emph{iii})
perform temporal cross-correlation of $\bar{\bq}(\bmm)$ and
$\partial_{tt}\bar{\bu}(\bmm)$. The gradient with respect to $\by$ in
equation~\eqref{eq:lagr_red_grady} corresponds to the data residual of
the augmented wavefield in equation~\eqref{eq:varproj}, plus a
relaxation term.

The fundamental gain in solving WRI$\ast$ instead of WRI lies in the
fact that the evaluation of the Lagrangian $\LWRId$ and its gradients
only require solutions of a conventional wave equation, opening up the
choice for time-domain solvers. The multipliers $\by$ can be stored in
memory whenever $\bmm$ and $\by$ are jointly optimized. Alternatively,
one might consider variable projection for $\by$. Both these approaches
are computationally expensive, and in this paper we will rely on a
simple approximation for $\by$, to be discussed in the next section.

\subsection{Dual variable
approximation}\label{dual-variable-approximation}

Equating the gradient with respect to $\by$ to zero, in
equation~\eqref{eq:lagr_red_grady}, provides a closed-form solution for
the optimal $\bar{\by}(\bmm)$, defined by the non-linear system:
\begin{equation}
\epsilon\dfrac{\bar{\by}(\bmm)}{\norm{\bar{\by}(\bmm)}}+\F\F^{\ast}\bar{\by}(\bmm)=\br,
\label{eq:WRIdual_yopt}
\end{equation}
 analogously to equation~\eqref{eq:yopt} for conventional WRI. Note that
a dualization of the data misfit term in the penalty method in
equation~\eqref{eq:pen} \citep[following, for
example,][]{chambolle2011first} would produce a Lagrangian
\begin{equation}
\LWRIdl(\bmm,\by)=-\dfrac{1}{2}\norm{\F^{\ast}\by}^2+\<\by,\br{}\>-\dfrac{\phantom{^2}\lambda^2}{2}\norm{\by}^2
\label{eq:WRIdual_yopt_lin}
\end{equation}
 quite similar to the Lagrangian in equation~\eqref{eq:lagr_red}. Not
surprisingly, the optimal $\bar{\by}_{\lambda}(\bmm)$ for
equation~\eqref{eq:WRIdual_yopt_lin} solves the same linear system in
equation~\eqref{eq:yopt} for WRI, and the reduced problem is equivalent
to WRI, since
$\LWRIdl(\bmm,\bar{\by}_{\lambda}(\bmm))=\JWRI(\bmm)/\lambda^2$.

Equation~\eqref{eq:WRIdual_yopt} (or its linear
variant~\ref{eq:WRIdual_yopt_lin}), can be solved iteratively. However,
each evaluation of equation~\eqref{eq:WRIdual_yopt} amounts to two wave
equation solutions and many iterations may be required. As a result, we
propose a cheap approximation for $\bar{\by}(\bmm)$ corresponding to the
scaled data residual
$\bar{\by}(\bmm)\approx\tilde{\by}(\bmm)=\alpha\br$, with an unknown
scaling factor $\alpha$. Given that equation~\eqref{eq:lagr_red} is
(strictly) concave with respect to $\by$, the optimal scaling
$\tilde{\alpha}(\bmm)=\arg\max_{\alpha}\LWRId(\bmm,\alpha\br)$ is
uniquely solved by
\begin{equation}
\begin{split}
& \tilde{y}=\tilde{\alpha}(\bmm)\br,\\
& \tilde{\alpha}(\bmm)=\left\{
\begin{array}{ll}
\dfrac{\norm{\br}(\norm{\br}-\epsilon)}{\norm{\F^{\ast}\br}^2}, & \norm{\br}\ge\epsilon,\\
0, & \mathrm{otherwise}.\\
\end{array}
\right.
\end{split}
\label{eq:alpha}
\end{equation}
 There are no significant extra costs associated with this computation,
as it can be obtained as a byproduct of the calculations involved in the
evaluation of the Lagrangian functional.
The scaled residual as an
approximation improves on the approach taken in \citet{wang2016full},
previously discussed, thanks to the optimal scaling. More interestingly,
owing to its simplicity, it allows analytical differentiation, as shown
in the next section.

\subsection{A new objective based on the dual
formulation}\label{a-new-objective-based-on-the-dual-formulation}

The approximation for $\by$ introduced in equation~\eqref{eq:alpha}
results in the reduced objective
\begin{equation}
\LWRIdg(\bmm)\defeq\LWRId(\bmm,\tilde{\by}(\bmm)),
\label{eq:newobj}
\end{equation}
 and its gradient with respect to $\bmm$ is
\begin{equation}
\nabla_{\bmm}\LWRIdg = \nabla_{\bmm}\LWRId(\bmm,\tilde{\by}(\bmm))-\tilde{\alpha}J[\bmm,\bq]^{\ast}\nabla_{\by}\LWRId(\bmm,\tilde{\by}(\bmm)),
\label{eq:WRIdg_gradm_corr}
\end{equation}
 where $\nabla_{\bmm}\LWRId$ and $\nabla_{\by}\LWRId$ are given in
equations~\eqref{eq:lagr_red_gradm} and~\eqref{eq:lagr_red_grady}. For
simplicity, we will still refer to the reduced problem as WRI$\ast$.
Note that the computation of equation~\eqref{eq:WRIdg_gradm_corr} does
not require the differentiation of $\tilde{\alpha}(\bmm)$ with respect
to $\bmm$, as its expression has been obtained from variable projection.
Neglecting the second term in equation~\eqref{eq:WRIdg_gradm_corr}
\citep[analogously to][]{wang2016full} saves some computation but, as
also observed in practice, it has a
detrimental effect on the convergence of nonlinear optimization methods,
such as l-BFGS
\citep{nocedal2006numerical}. The evaluation and gradient calculation of
the WRI$\ast$ objective~\eqref{eq:newobj} is schematically reported in
algorithm~\ref{alg}, and can be used in combination with any nonlinear
optimization solver.

\begin{scholmdAlgorithm}
\textbf{Notation}:\\\hspace*{0.333em}\hspace*{0.333em}\hspace*{0.333em}\hspace*{0.333em}\hspace*{0.333em}\hspace*{0.333em}$R$:~receiver~restriction~operator\\\hspace*{0.333em}\hspace*{0.333em}\hspace*{0.333em}\hspace*{0.333em}\hspace*{0.333em}\hspace*{0.333em}$\A$:~wave~equation~operator\\\textbf{Function}:~\texttt{objective\_WRI}\\\textbf{Input}:\\\hspace*{0.333em}\hspace*{0.333em}\hspace*{0.333em}\hspace*{0.333em}\hspace*{0.333em}\hspace*{0.333em}$\epsilon$:~fixed~data~tolerance\\\hspace*{0.333em}\hspace*{0.333em}\hspace*{0.333em}\hspace*{0.333em}\hspace*{0.333em}\hspace*{0.333em}$\mathbf{d}$:~observed~data\\\hspace*{0.333em}\hspace*{0.333em}\hspace*{0.333em}\hspace*{0.333em}\hspace*{0.333em}\hspace*{0.333em}$\mathbf{q}$:~source~position/wavelet\\\hspace*{0.333em}\hspace*{0.333em}\hspace*{0.333em}\hspace*{0.333em}\hspace*{0.333em}\hspace*{0.333em}$\bmm$:~model\\\textbf{Algorithm}:\\1.~~~~Forward~wavefield~$\bu=\A^{-1}\bq$~and~residual~$\mathbf{r}=\bd-\R\bu$\\2.~~~~Backward~wavefield~$\tilde{\bq}=\A^{-*}\R^{*}\br$\\3.~~~~Compute~$\alpha=\alpha(\norm{\mathbf{r}},\norm{\tilde{\bq}},\epsilon)$,~equation~\eqref{eq:alpha}\\4.~~~~Objective~value~$L=-\alpha^2\norm{\tilde{\bq}}^2/2+\alpha\norm{\mathbf{r}}^2-|\alpha|\epsilon\norm{\mathbf{r}}$,~equation~\eqref{eq:newobj}\\5.~~~~Augmented~wavefield~$\tilde{\bu}=\A^{-1}(\bq+\alpha\tilde{\bq})$~and~residual~$\tilde{\mathbf{r}}=\bd-R\tilde{\bu}$\\6.~~~~Gradient~$\mathbf{g}_{\bmm,1}(\bx)=-\alpha\sum_t\partial_{tt}\tilde{\bu}(\bx,t)\tilde{\bq}(\bx,t)$,~equation~\eqref{eq:lagr_red_gradm}\\7.~~~~Gradient~$\mathbf{g}_{\by}=\tilde{\mathbf{r}}-\mathrm{sign}(\alpha)\,\epsilon\,\mathbf{r}/\norm{\mathbf{r}}$,~equation~\eqref{eq:lagr_red_grady}\\8.~~~~Backpropagated~gradient~$\bv=\A^{-*}\R^{*}\mathbf{g}_{\by}$\\9.~~~~Gradient~correction~$\mathbf{g}_{\bmm,2}=-\alpha\sum_t\partial_{tt}\bu(\bx,t)\bv(\bx,t)$\\10.~~Corrected~gradient~$\mathbf{g}_{\bmm}=\mathbf{g}_{\bmm,1}+\mathbf{g}_{\bmm,2}$,~equation~\eqref{eq:WRIdg_gradm_corr}\\\textbf{Output:}~$L$,$\mathbf{g}_{\bmm}$
\caption{WRI$\ast$ objective evaluation and gradient computation.
}\label{alg}
\end{scholmdAlgorithm}

\subsection{\texorpdfstring{Source-dependent metrics for the wave
equation
misfit\label{sec:srcweight}}{Source-dependent metrics for the wave equation misfit}}\label{source-dependent-metrics-for-the-wave-equation-misfit}

In equation~\eqref{eq:denoise}, we can exploit the knowledge about the
location and distribution of the physical source $\bq$, and design
metrics for the wave equation misfit that are more suited for the
problem at hand. As previously discussed, \citet{huang2018volume} choose
a weighted least-squares norm that penalizes functions with a support
distributed far from the point source location.
Some other recent work by
\citet{symes2020waveform}, demonstrates that WRI is susceptible to local
minima when no source-weighting is
applied. Here, we then select a penalty
grid function $\bw(\bx)=\sqrt{|\bx-\bx_{\mathrm{s}}|^2+h^2}/h$ dependent
on a parameter $h$ tuned to be a fraction of a reference wavelength.
With this modification, the Lagrangian in equation~\eqref{eq:lagr_red}
becomes
\begin{equation}
\LWRId(\bmm,\by)=-\dfrac{1}{2}\norm{\F^{\ast}\by}_{\Sigma^{-1}}^2+\<\by,\br\>-\epsilon\norm{\by},
\label{eq:srcweight}
\end{equation}
 and only minor corrections are needed for related gradient expressions
and dual variable approximation. In the experiments in the next section,
we tacitly assume this type of source-dependent weighting. Note that
source weighting cannot be trivially incorporated in FWI, since it is
based on the exact solution of the wave equation.
Beside the theoretical considerations
in \citet{symes2020waveform}, source weighting is beneficial to enhance
the resolution for portions of the model located far from the source
position. We point out that the hyperparameter $h$ should be small
enough to promote the focusing of the additional source term
$\F^{\ast}\by$ in
equation~\eqref{eq:srcweight}.

Other potentially interesting modifications of the WRI$\ast$objective in
equation~\eqref{eq:lagr_red} involve the $\ell_1$ norm,
$\norm{\cdot}_1$, or weighted versions thereof, to measure the wave
equation error. The basic rationale is to impose sparsity on the spatial
distribution of the extended source. We refer to \citet{sharan2018fast}
for a microseismic application. Following \citet{sharan2018fast}, we
further suggest the so-called elastic net regularization obtained by
replacing the $\ell_1$ norm with a relaxation that involves the
least-squares norm: $\norm{\cdot}_1+\norm{\cdot}_2^2/(2\mu^2)$. This
relaxation is key in obtaining an analytical expression for the dual
objective, analogously to equation~\eqref{eq:lagr_red}. In this paper,
however, we defer the $\ell_1$ formulation to future work, and stick to
the objective~\eqref{eq:lagr_red}.

\subsection{\texorpdfstring{Complexity of
WRI$\ast$}{Complexity of WRI\textbackslash{}ast}}\label{complexity-of-wriast}

Compared to FWI, evaluating and differentiating~\eqref{eq:newobj}
requires the computation of twice as many solutions of the wave
equation, respectively. Despite the increased cost, this still compares
favorably to conventional WRI, since we avoid the need to solve for an
augmented wave equation and frequency domain solvers. Moreover, as it
will be demonstrated with numerical examples, WRI$\ast$ retains the
robustness against local minima characteristic of classical WRI, despite
the approximations discussed in the previous sections.

\subsection{Beyond the dual variable
approximation}\label{beyond-the-dual-variable-approximation}

The approximation for the dual
variable introduced in equation~\eqref{eq:alpha} is motivated by
computational reasons and the need for analytical expressions for the
objective gradient, here available precisely because of its simple
expression. Naturally, a more accurate approximation of the optimal
solution $\bar{\by}(\bmm)$ in equation~\eqref{eq:WRIdual_yopt} can be
attained by iteratively solving the non-linear system in
equation~\eqref{eq:WRIdual_yopt}. If we denote by $\tilde{\by}^n(\bmm)$
the estimate obtained after $n$ iterations of a given solver for the
equation~\eqref{eq:WRIdual_yopt}, we can define the corresponding
reduced objective by
$\LWRIdg^n(\bmm)\defeq\LWRId(\bmm,\tilde{\by}^n(\bmm))$, analogously to
equation~\eqref{eq:newobj}. Clearly, WRI$\ast$ corresponds to $n=1$. For
$n>1$, differentiating $\LWRIdg^n$ will likely require automatic
differentiation \citep{griewank2008evaluating}. Note that one can
neglect the dependency of the augmented variable with respect to $\bmm$
with adaptively accurate solves, as demonstrated in \citet{van20143d},
so that an exact gradient computation is not strictly needed for
convergence. Recently, however, it has been shown that rigorous gradient
calculations lead to ``super-efficient'' estimators for the optimal
gradient \citep{ablin2020superefficiency}. Investigating this line of
research is not within the scope of this paper and is left to future
studies.

\section{Results}\label{results}

We test the capabilities of the WRI$\ast$ inversion method with
synthetic examples.

We start by comparing WRI$\ast$ with classical WRI. We remind that
WRI$\ast$ and WRI are equivalent only when the dual variable is solved
exactly, as in equation~\eqref{eq:WRIdual_yopt} (with the caveat
described in that section). In other words, we will test the effect of
the dual variable approximation proposed in equation~\eqref{eq:alpha}.
Note that this set of experiments will be carried out in the frequency
domain, since traditional WRI is therein framed more conveniently from a
computational point of view.

Other experiments mainly aim at the empirical demonstrations of the
following claims:

\begin{itemize}
\itemsep1pt\parskip0pt\parsep0pt
\item
  WRI$\ast$ is more robust than FWI with respect to local minima;
\item
  WRI$\ast$ is more robust than FWI with respect to modeling
  inaccuracies.
\end{itemize}

With modeling inaccuracies, we refer to erroneous assumptions about some
of the physical parameters that will be kept fixed during inversion, in
the context of TTI modeling (tilted transverse isotropy) and inversion.

Finally, we present a small-sized 3D
problem with a comparison of initial gradients of WRI$\ast$ and FWI, as
a testament for potential full-sized
applications.

\subsection{\texorpdfstring{WRI vs
WRI$\ast$}{WRI vs WRI\textbackslash{}ast}}\label{wri-vs-wriast}

The aim of this experiment is a qualitative assessment of the dependency
of the WRI$\ast$ scheme on the hyperparameter $\epsilon$, appearing in
equation~\eqref{eq:denoise} (which denotes a given data noise
tolerance), and the comparison with WRI. We emphasize that one of the
main differences between WRI$\ast$ and classical WRI is a suboptimal ---
but computationally convenient --- approximation of the slack variables
$\by$ described in~\eqref{eq:alpha} (while the optimal $\by$ makes
WRI$\ast$ and WRI substantially equivalent). The scope of the example
described in this section is therefore twofold:

\begin{itemize}
\itemsep1pt\parskip0pt\parsep0pt
\item
  determine the effect of $\epsilon$ on the WRI$\ast$ inversion result;
\item
  determine the effect of the dual variable approximation in
  equation~\eqref{eq:alpha} by comparing WRI$\ast$ with traditional WRI
  (based on the optimal solution for $\by$ via variable projection).
\end{itemize}

The classical WRI formulation is defined by the objective functional in
equation~\eqref{eq:WRIdual_yopt_lin}, where $\by$ is obtained
by~\eqref{eq:yopt}. Note that the experiment of this section will be
carried out in the frequency domain, where WRI is more conveniently
formulated.

\begin{figure}
\centering
\subfloat[\label{fig:marm-true}]{\includegraphics[width=1.000\hsize]{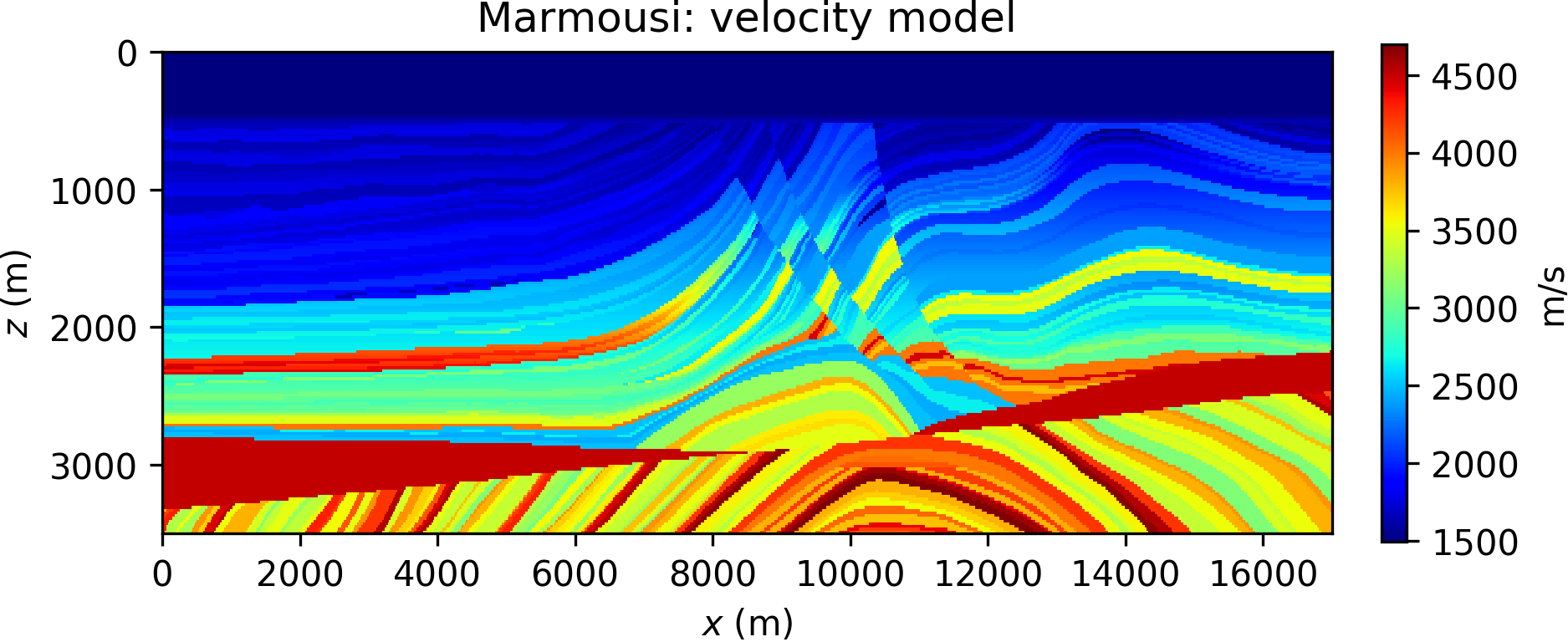}}
\\
\subfloat[\label{fig:marm-bg}]{\includegraphics[width=1.000\hsize]{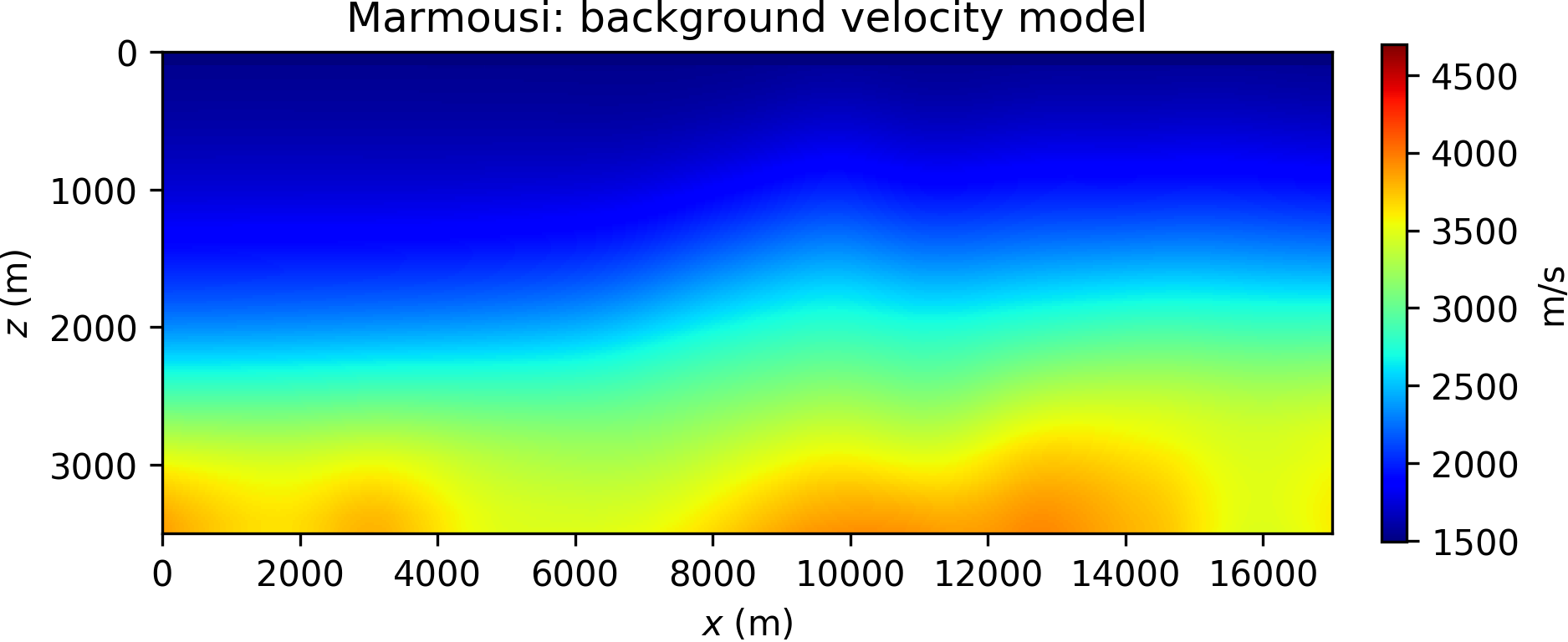}}
\caption{Marmousi model: (a) true velocity model, (b) background model.
We invert single-frequency data components with a multiscale approach
\citep{bunksFWI} from low to high frequencies, and run 10 iterations of
the l-BFGS optimizer per frequency. We compare FWI, WRI$\ast$ with
different values for $\epsilon$, and conventional WRI (with a fixed
choice for the weighting parameter $\lambda$ in
equation~\ref{eq:WRIdual_yopt_lin}). }\label{fig:marm}
\end{figure}

We consider the well-known Marmousi model, presented in
Figure~\ref{fig:marm-true}. Data are directly generated in the frequency
domain with components ranging from 3 Hz to 14 Hz, with no additional
noise (Figure \#fig:marmdata). We set 100 sources evenly spaced and
receivers densely sampled along the surface.

\begin{figure}
\centering
\subfloat[\label{fig:marminv-fwi}]{\includegraphics[width=0.500\hsize]{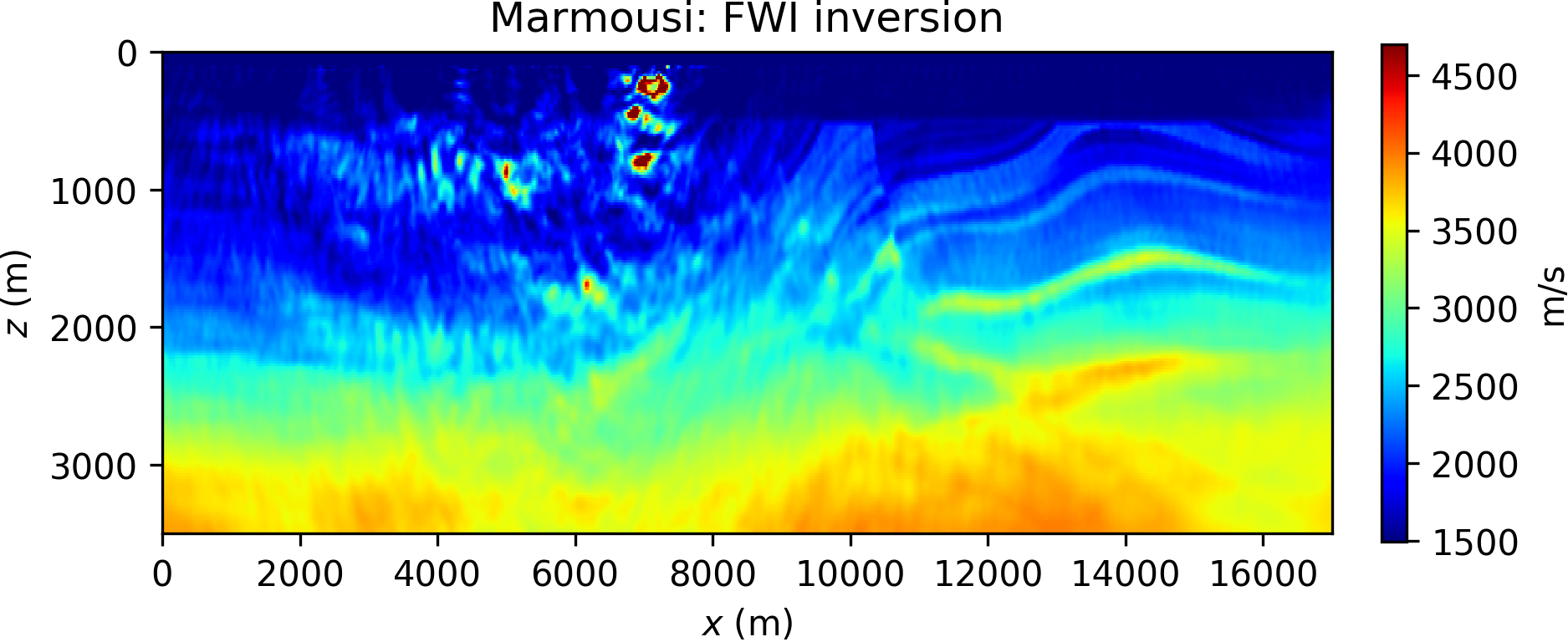}}
\\
\subfloat[\label{fig:marminv-wri1}]{\includegraphics[width=0.500\hsize]{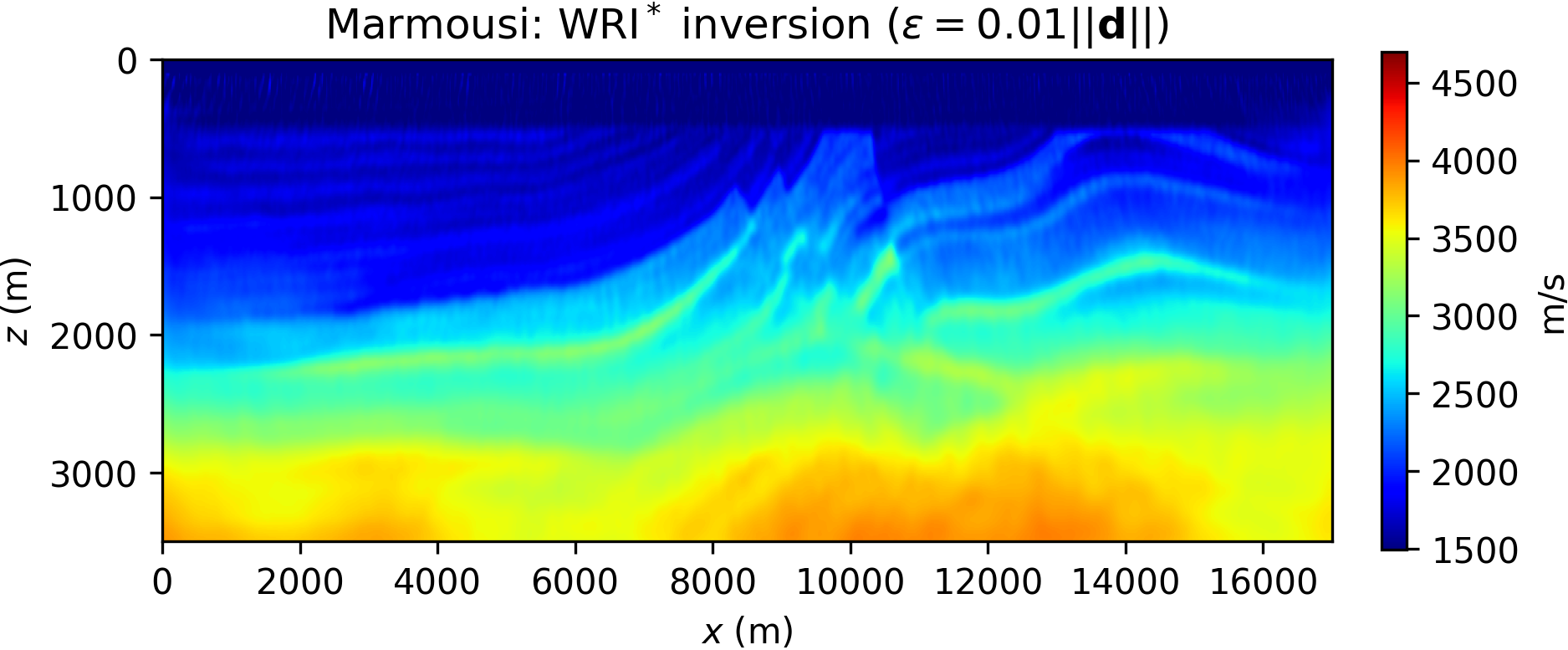}}
\\
\subfloat[\label{fig:marminv-wri2}]{\includegraphics[width=0.500\hsize]{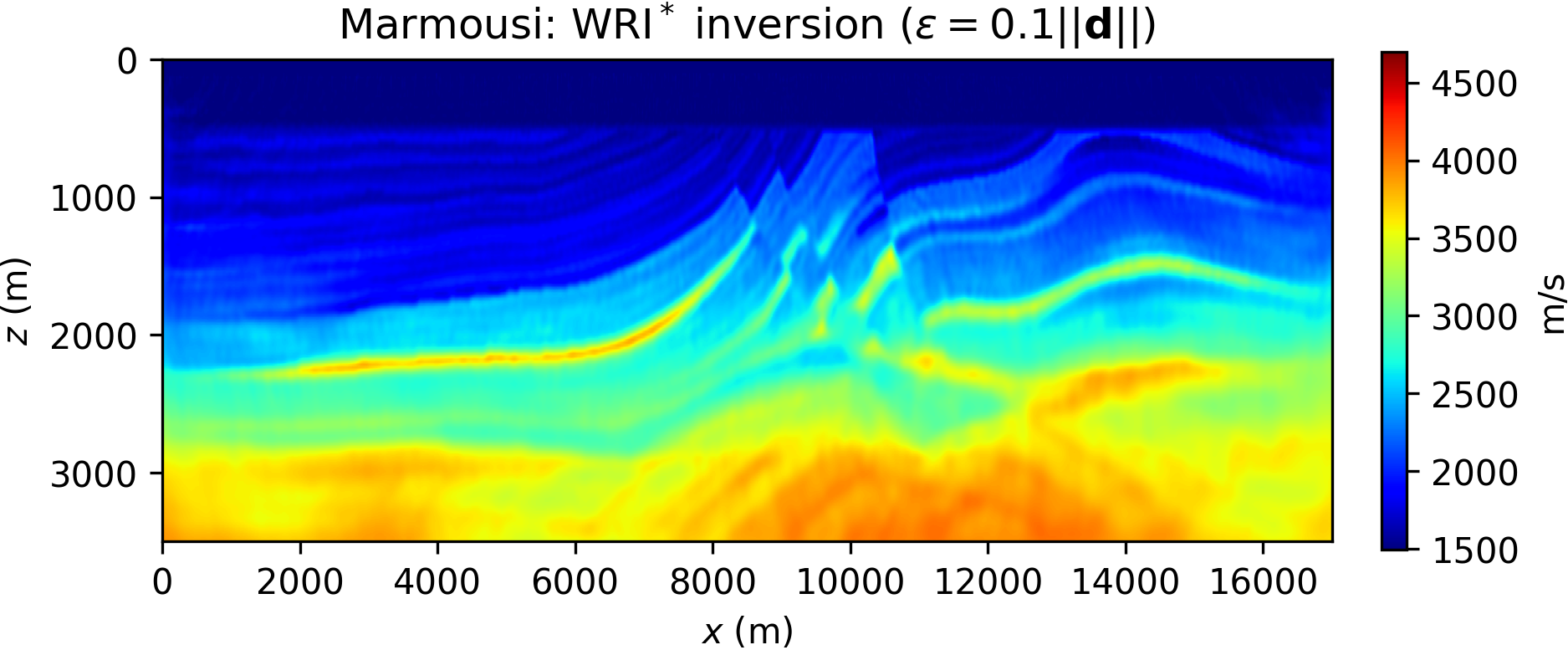}}
\\
\subfloat[\label{fig:marminv-wril}]{\includegraphics[width=0.500\hsize]{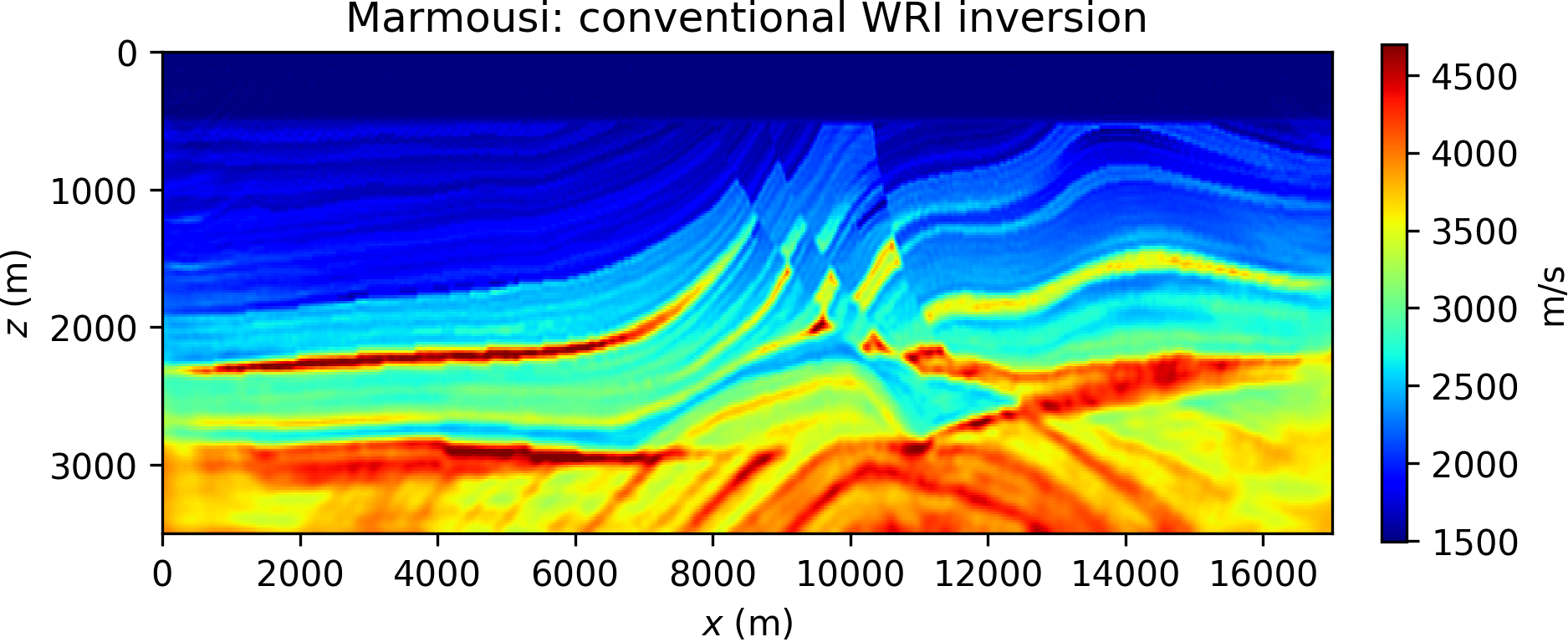}}
\\
\subfloat[\label{fig:marminv-fwiwri}]{\includegraphics[width=0.500\hsize]{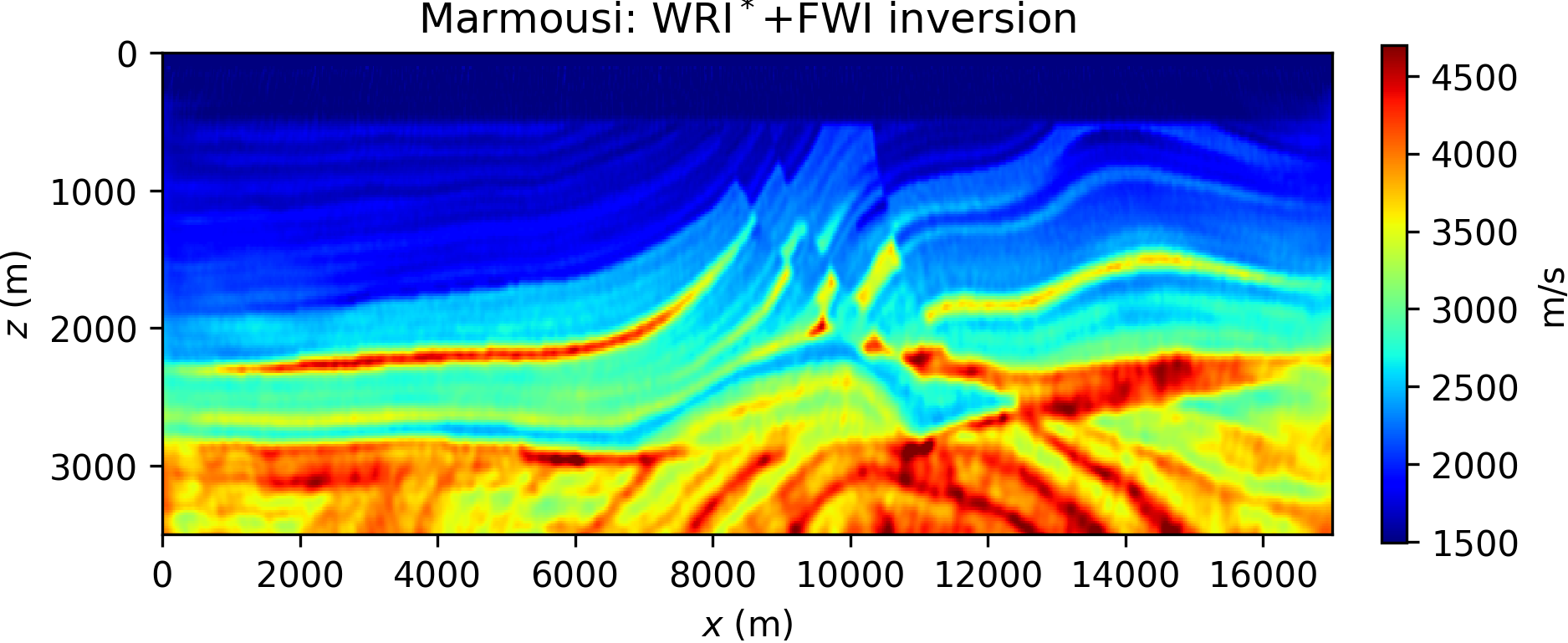}}
\caption{Marmousi inversion result for: (a) FWI, (b) WRI$\ast$
($\epsilon=0.01\norm{\bd}$), (c) WRI$\ast$ ($\epsilon=0.1\norm{\bd}$),
(d) conventional WRI, (e) sequential inversion: WRI$\ast$
($\epsilon=0.01\norm{\bd}$) followed by FWI with Gauss-Newton
preconditioning.}\label{fig:marminv}
\end{figure}

Comparing the results in Figure~\ref{fig:marminv}, it is apparent that
FWI converges to a local minimum, while both WRI$\ast$ and conventional
WRI produce plausible models. WRI$\ast$ does not achieve the same
resolution of the WRI inversion, although we observe that increasing
values for $\epsilon$ translate to higher model resolution
(cf.~Figures~\ref{fig:marminv-wri1} and~\ref{fig:marminv-wri2}). This
behavior can be understood in the following qualitative terms: lower
values for $\epsilon$ promote a wavefield solution that fits data rather
than the wave equation (see, for instance, equation~\ref{eq:denoise}).
Relaxing the wave equation is
beneficial to avoid local minima, as advocated in the original WRI
method, but it translates to inaccurate model resolution. Note that, in
WRI, the $\lambda$ parameter appearing in equation~\eqref{eq:WRI} plays
the same role as $\epsilon$.

Despite the similarities between WRI
and WRI$\ast$, the WRI result is generally less affected by the choice
of the relaxation parameter $\lambda$ compared to WRI$\ast$ with
$\epsilon$ (see Figure~\ref{fig:marminv-wril}). This behavior is
dictated by the sub-optimal dual variable estimate in WRI$\ast$. In our
experience, however, WRI$\ast$ conserves the same ability of WRI to
circumvent local minima, and the gap between the two methods can be
reduced by either quasi-Newton preconditioning or devising a sequential
WRI$\ast$-FWI scheme where the WRI$\ast$ result is sharpened by FWI
(which will also help in saving computations once local minima are
avoided). This approach is also suggested in \citet{warnerAWI2}. An
example of the joint WRI$\ast$-FWI is depicted in
Figure~\ref{fig:marminv-fwiwri}, highlighting a qualitatively similar
result to conventional WRI in Figure~\ref{fig:marminv-wril}. Note that
this hybrid scheme might lose the benefit of WRI$\ast$ in dealing with
inexact modeling assumptions. In that case, we might replace the
least-squares misfit in FWI with more robust objectives as proposed in
\citet{herrmann2013domain}.

\begin{figure}
\centering
\subfloat[\label{fig:marm-rtm0}]{\includegraphics[width=0.800\hsize]{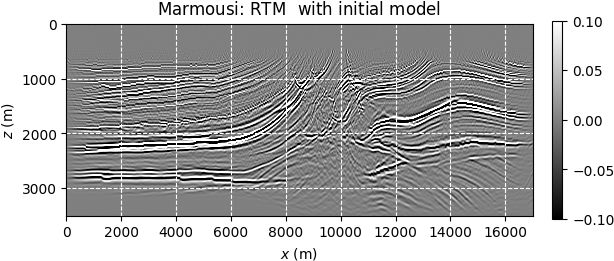}}
\\
\subfloat[\label{fig:marm-rtm1}]{\includegraphics[width=0.800\hsize]{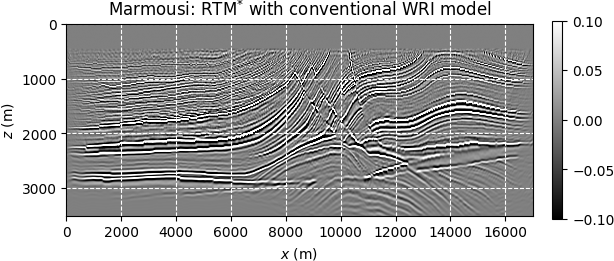}}
\\
\subfloat[\label{fig:marm-rtm2}]{\includegraphics[width=0.800\hsize]{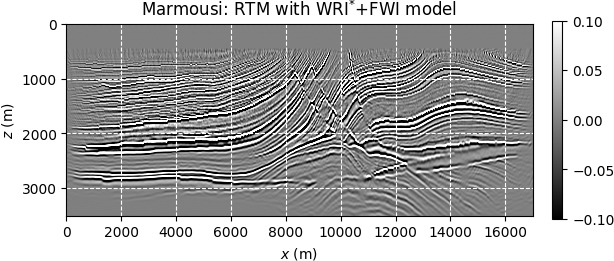}}
\caption{Reverse time migrations (normalized for display purposes) for
the Marmousi model obtained from the following velocity models: (a)
starting model (Figure~\ref{fig:marm-bg}), (b) conventional WRI
(Figure~\ref{fig:marminv-wril}), (c) combined WRI${\ast}$/FWI strategy
(Figure~\ref{fig:marminv-fwiwri}). A grid is overlayed in white to aid the comparison.}\label{fig:marm-rtm}
\end{figure}

Finally, in Figure~\ref{fig:marm-rtm}
we compare reverse time migration \citep[RTM,][]{baysal1983reverse}
obtained from different inversion results. For this purpose,
high-frequency time-domain synthetic data is generated with the same
source-receiver acquisition parameters used in the inversion phase, but
with a Ricker source wavelet of 20 Hz peak frequency. The comparison is
made with RTM obtained from the initial background, conventional WRI,
and hybrid WRI$\ast$/FWI inversion results. All these results improve on
the starting RTM image, with higher resolution, more continuous, and
more accurately positioned reflectors (especially in the central portion
of the model). The migration quality clearly improves with the
resolution of the inverted model, with deeper reflectors better resolved
by conventional WRI and hybrid WRI${\ast}$/FWI
strategies.

\subsection{\texorpdfstring{FWI vs
WRI$\ast$}{FWI vs WRI\textbackslash{}ast}}\label{fwi-vs-wriast}

\subsubsection{Low-velocity lens}\label{low-velocity-lens}

\begin{figure}
\centering
\includegraphics[width=0.750\hsize]{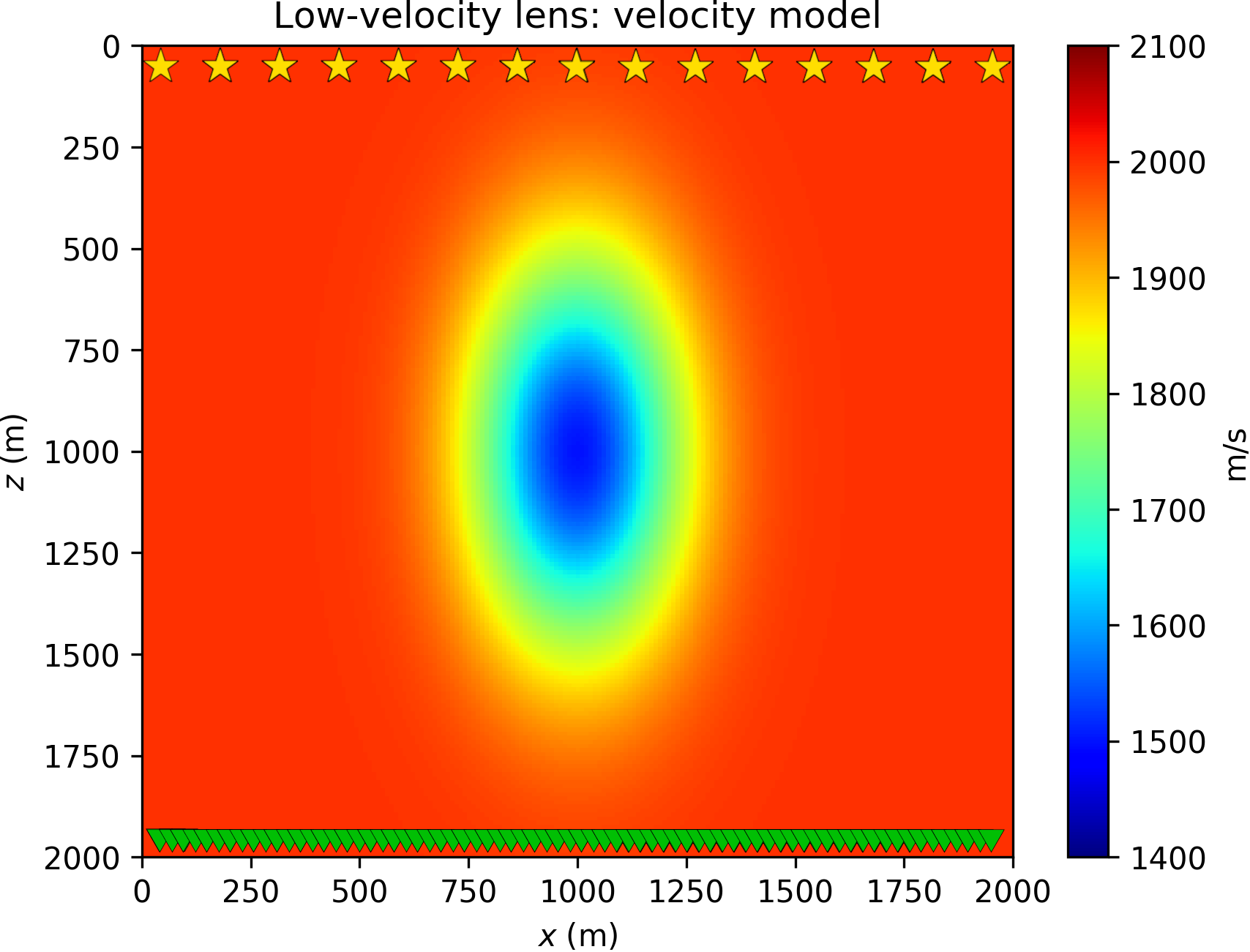}
\caption{Low-velocity lens model. Sources are indicated by stars, and
receivers by triangles. In this experiment, we run 20 iterations of the
l-BFGS optimizer \citep{nocedal2006numerical}. Due to cycle skipping,
FWI converges to a local minimum, shown in Figure~\ref{fig:lensinv-fwi}.
We stress the fact both WRI or WRI$\ast$ need be used in combination
with source weighting (equation~\ref{eq:srcweight}) to converge to the
solution \citep[as also documented
in][]{huang2018volume, symes2020waveform, symes2020wavefield}. The
WRI$\ast$ result (obtained with data tolerance $\epsilon=0$) is depicted
in Figure~\ref{fig:lensinv-twri}. We observe that WRI$\ast$ is clearly
able to reconstruct the anomaly and is not stuck to a local
minimum.}\label{fig:lens}
\end{figure}

This numerical experiment is geared towards a direct comparison of
WRI$\ast$ and FWI, and demonstrates the relative robustness of WRI$\ast$
against spurious modes. We consider the Gaussian lens model presented in
\citep{huang2018volume}, here reproduced in Figure~\ref{fig:lens}. The
problem consists of imaging a low-velocity anomaly of Gaussian shape
from transmission data, starting with an homogeneous model. Contrary to
the previous example, we run this experiment in time domain. The
synthetics are generated with a Ricker source wavelet of 10 Hz peak
frequency (corresponding to a spatial wavelength of 200 m), with no
artificial noise added. We model 15 sources, and gather data with
approximately 200 receivers (evenly spaced). The low-velocity zone
generates triplications and eventually causing cycle skipping.

\begin{figure}
\centering
\subfloat[\label{fig:lensinv-fwi}]{\includegraphics[width=0.750\hsize]{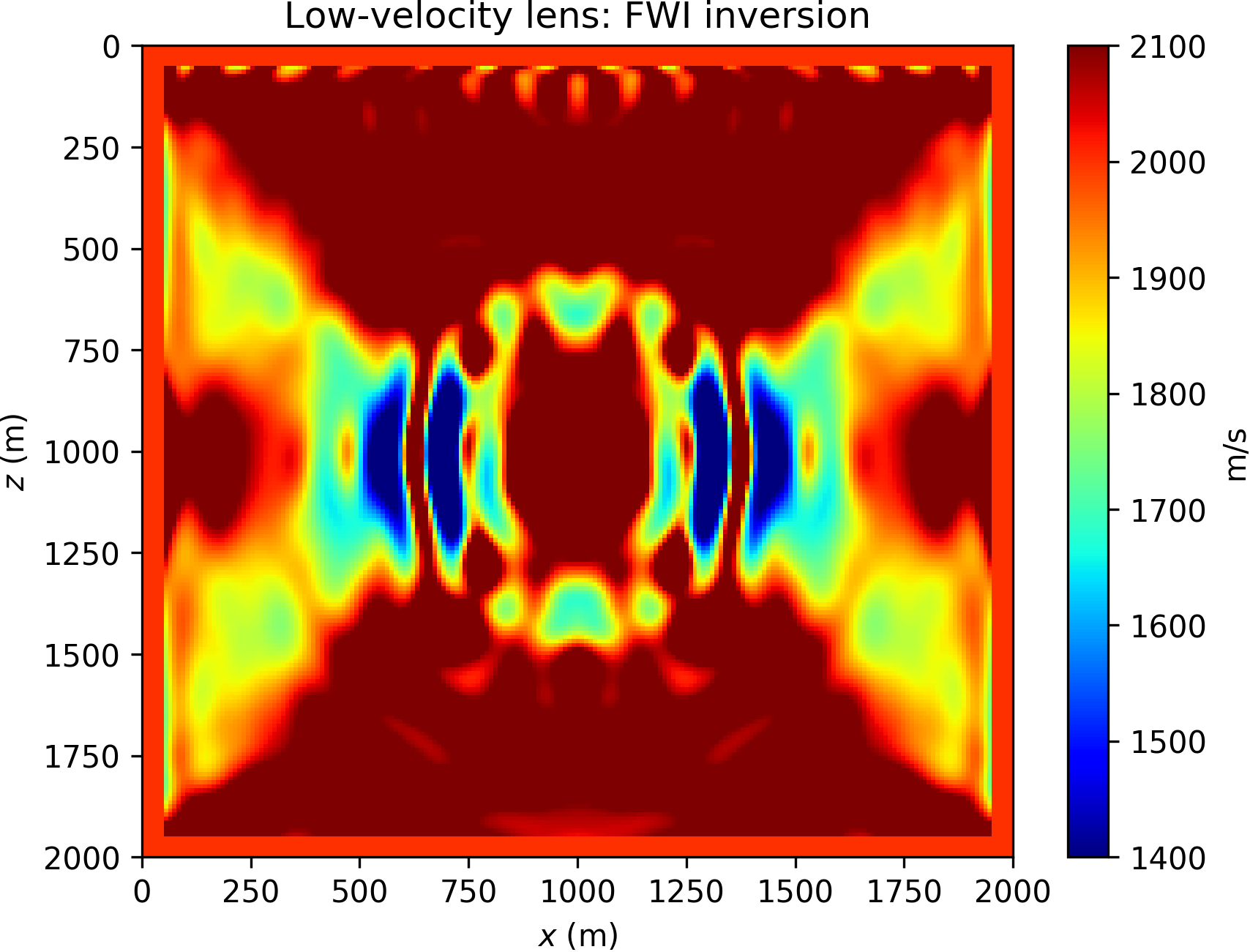}}
\\
\subfloat[\label{fig:lensinv-twri}]{\includegraphics[width=0.750\hsize]{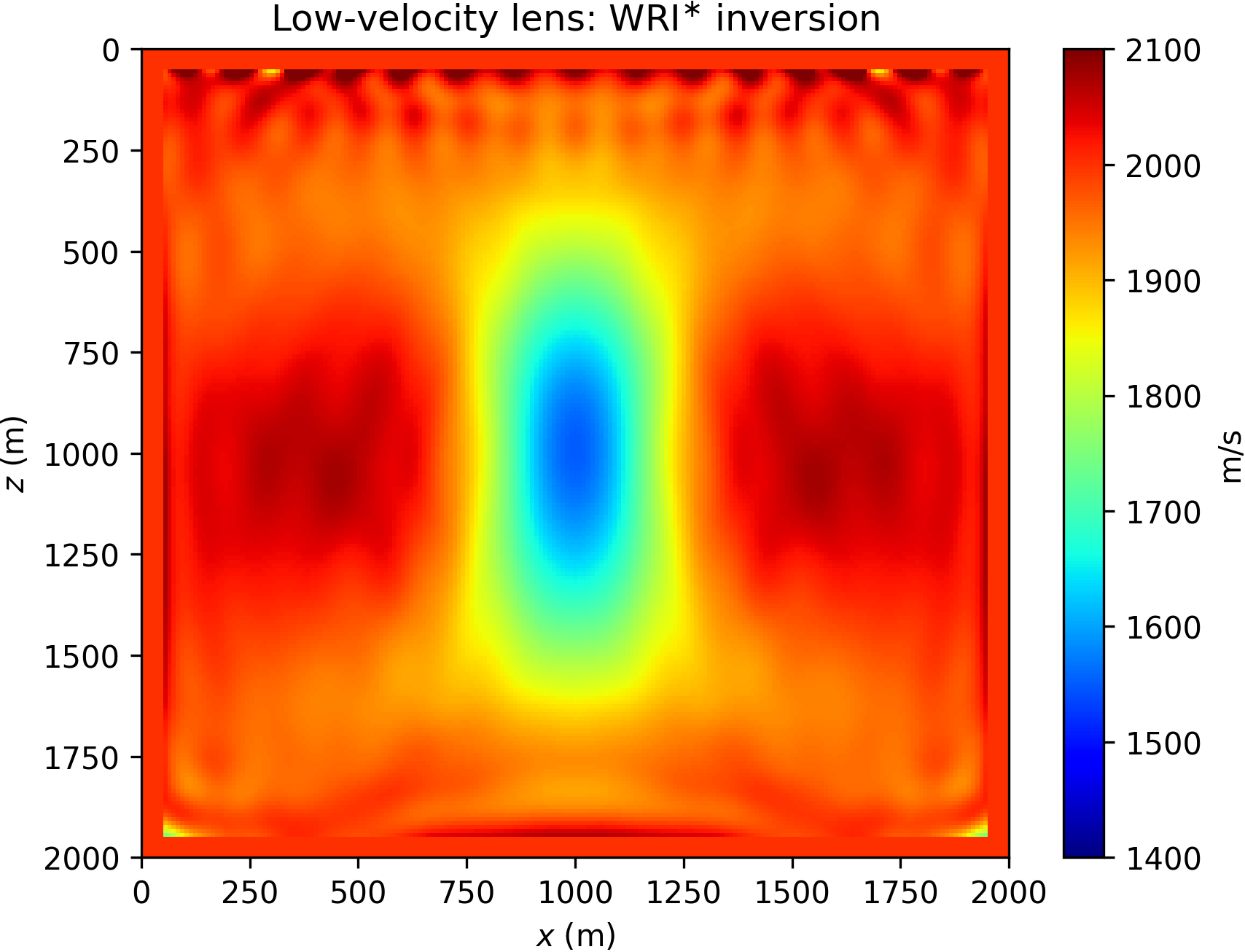}}
\caption{Inversion results for the Gaussian lens problem: (a) FWI
inversion result, (b) WRI$\ast$ inversion result.}\label{fig:lensinv}
\end{figure}

\subsubsection{BG Compass}\label{bg-compass}

We focus now not only on how WRI$\ast$ and FWI handle cycle skipping,
but also how inaccurate assumptions about the underlying physical model
of the data reflect on the inversion result. For example, what is the
effect of not accounting for the correct anisotropic model when data
come from anisotropic modeling? We show that WRI$\ast$, being based on a
relaxation of the wave equation, has the potential to mitigate erroneous
assumptions, while FWI will be more heavily affected. We therefore
compare the results of WRI$\ast$ and FWI on the BG Compass model
according to two distinct scenarios:

\begin{itemize}
\itemsep1pt\parskip0pt\parsep0pt
\item
  perfect modeling assumptions: both given data and modeled synthetics
  are obtained from the same physical model (in particular, acoustic
  modeling). The inversion parameter is the squared slowness;
\item
  inaccurate modeling assumptions: given data are produced with a
  certain TTI model (tilted transverse isotropy), while synthetics are
  modeled with an inaccurate TTI model. All the TTI parameters but the
  squared slowness are kept fixed during inversion.
\end{itemize}

As in the previous example, this set of experiments is carried out in
time domain. The data tolerance
(defined in equation~\ref{eq:denoise}) will be set to $\epsilon=0$ for
all the following experiments.

\paragraph{Inversion with correct modeling
assumptions}\label{inversion-with-correct-modeling-assumptions}

\begin{figure}
\centering
\subfloat[\label{fig:BGCmodelac-true}]{\includegraphics[width=1.000\hsize]{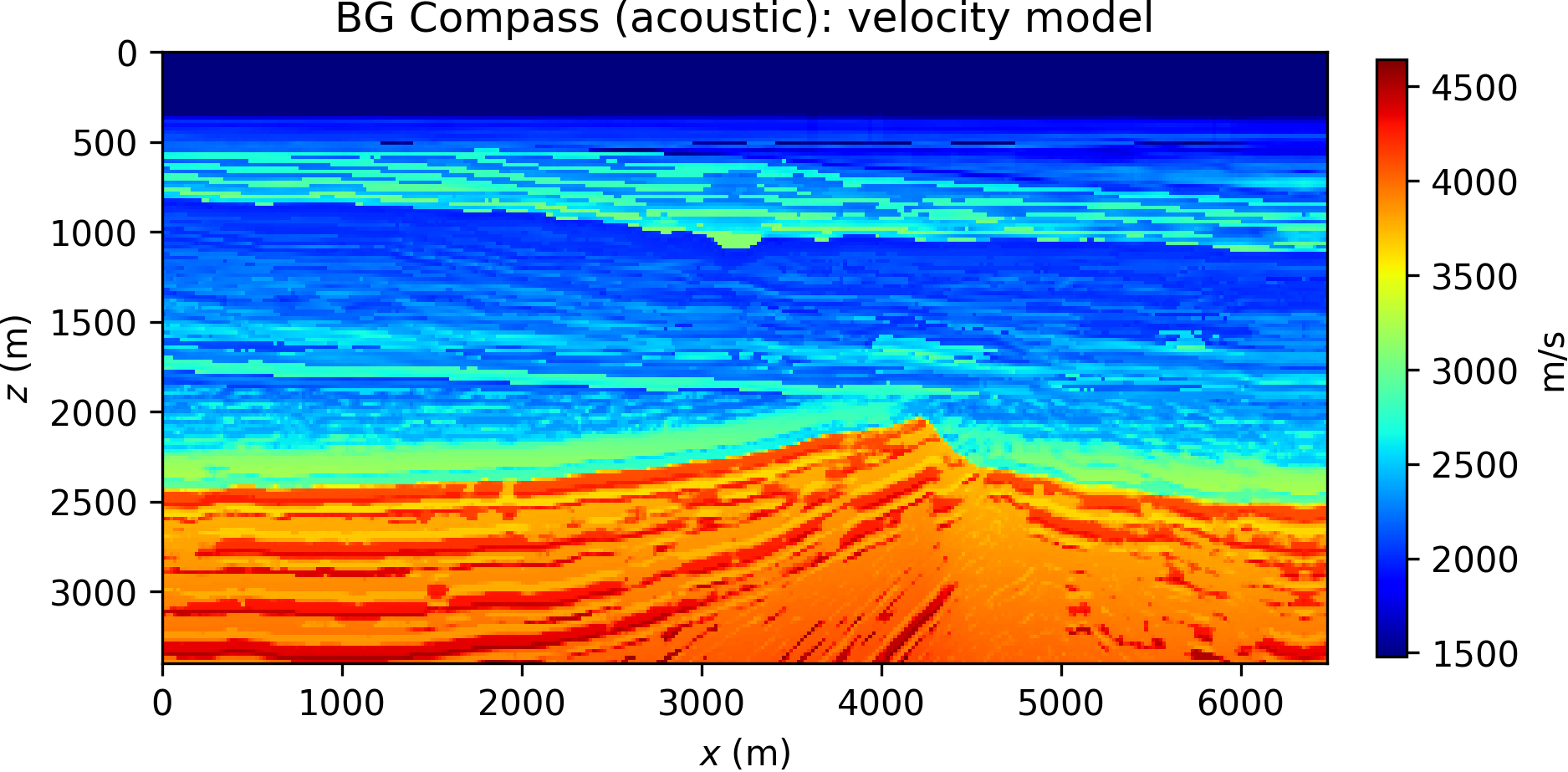}}
\\
\subfloat[\label{fig:BGCmodelac-bg}]{\includegraphics[width=1.000\hsize]{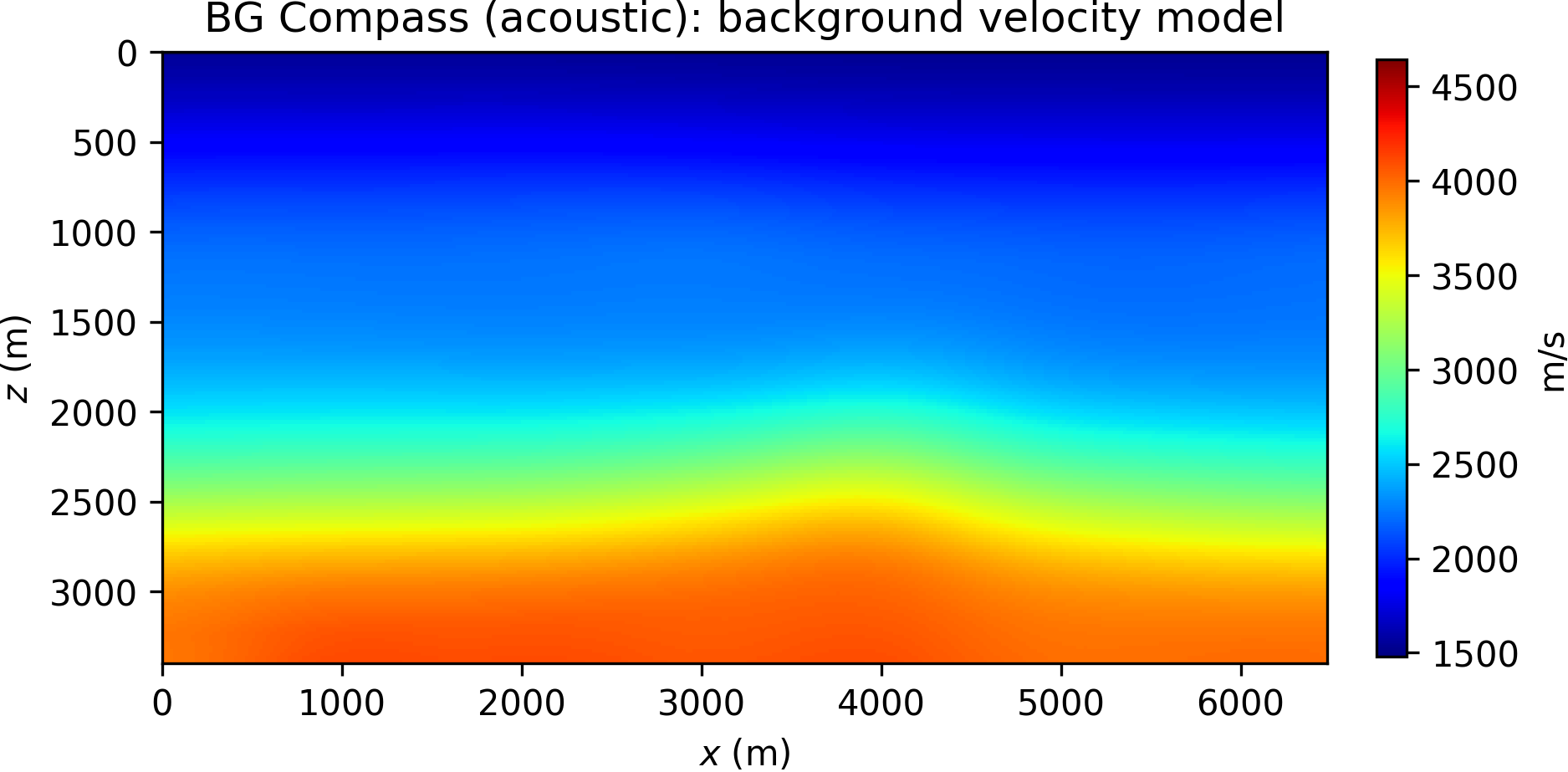}}
\caption{BG Compass, acoustic model: (a) velocity model, (b) background
model.}\label{fig:BGCmodelac}
\end{figure}

The acoustic (constant density) BG Compass model, here considered, is
representative of some of the complexities encounter in the field, and
leads to a notoriously difficult problem for FWI due to the velocity
kickback approximately located at the depth of 1 Km (see
Figure~\ref{fig:BGCmodelac}), delaying turning waves traveling back to
the surface. We emphasize that the BG
Compass here considered is made larger than the original version to exacerbate these
challenges.

\begin{figure}
\centering
\subfloat[\label{fig:BGCgrad-fwi}]{\includegraphics[width=1.000\hsize]{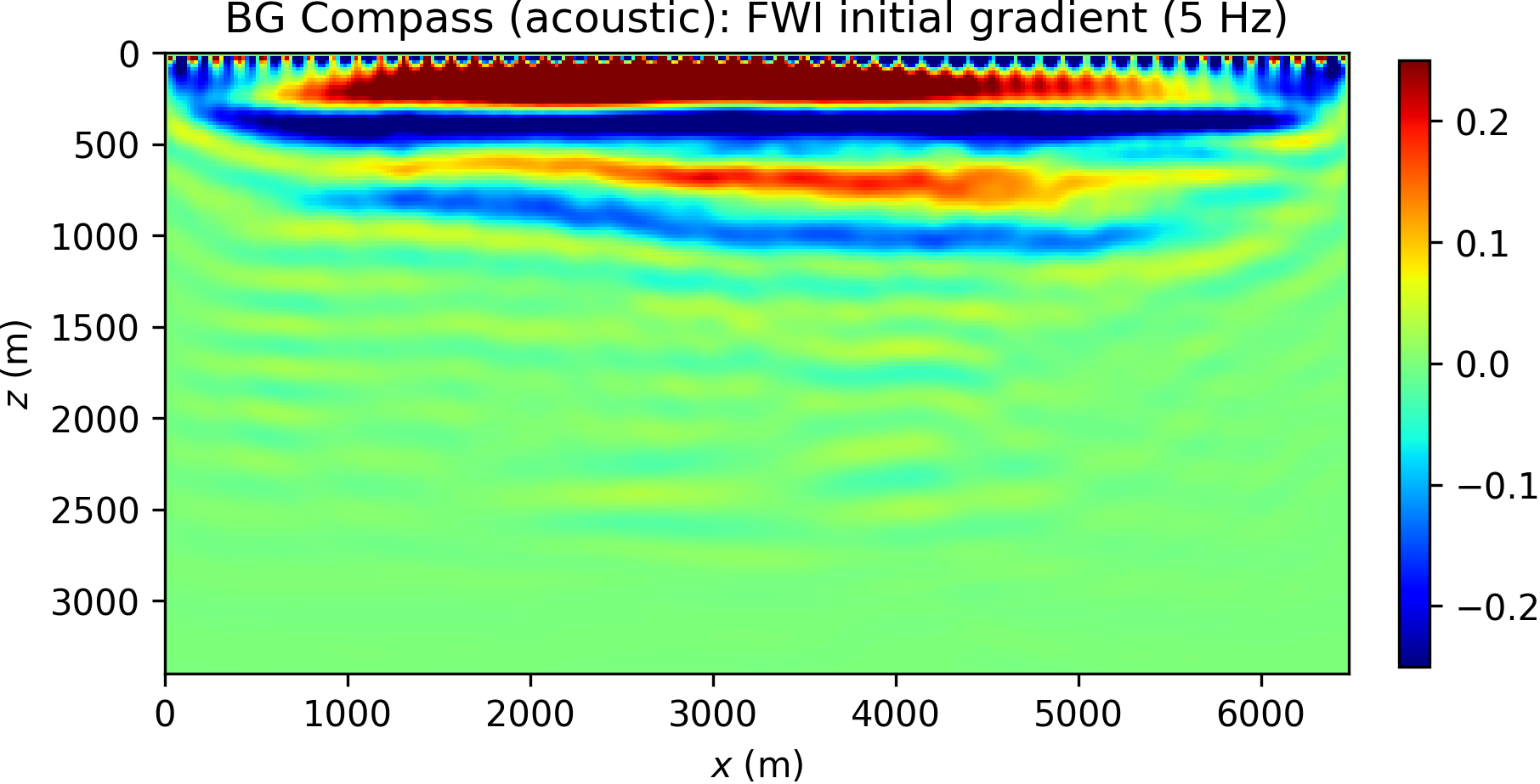}}
\\
\subfloat[\label{fig:BGCgrad-twri}]{\includegraphics[width=1.000\hsize]{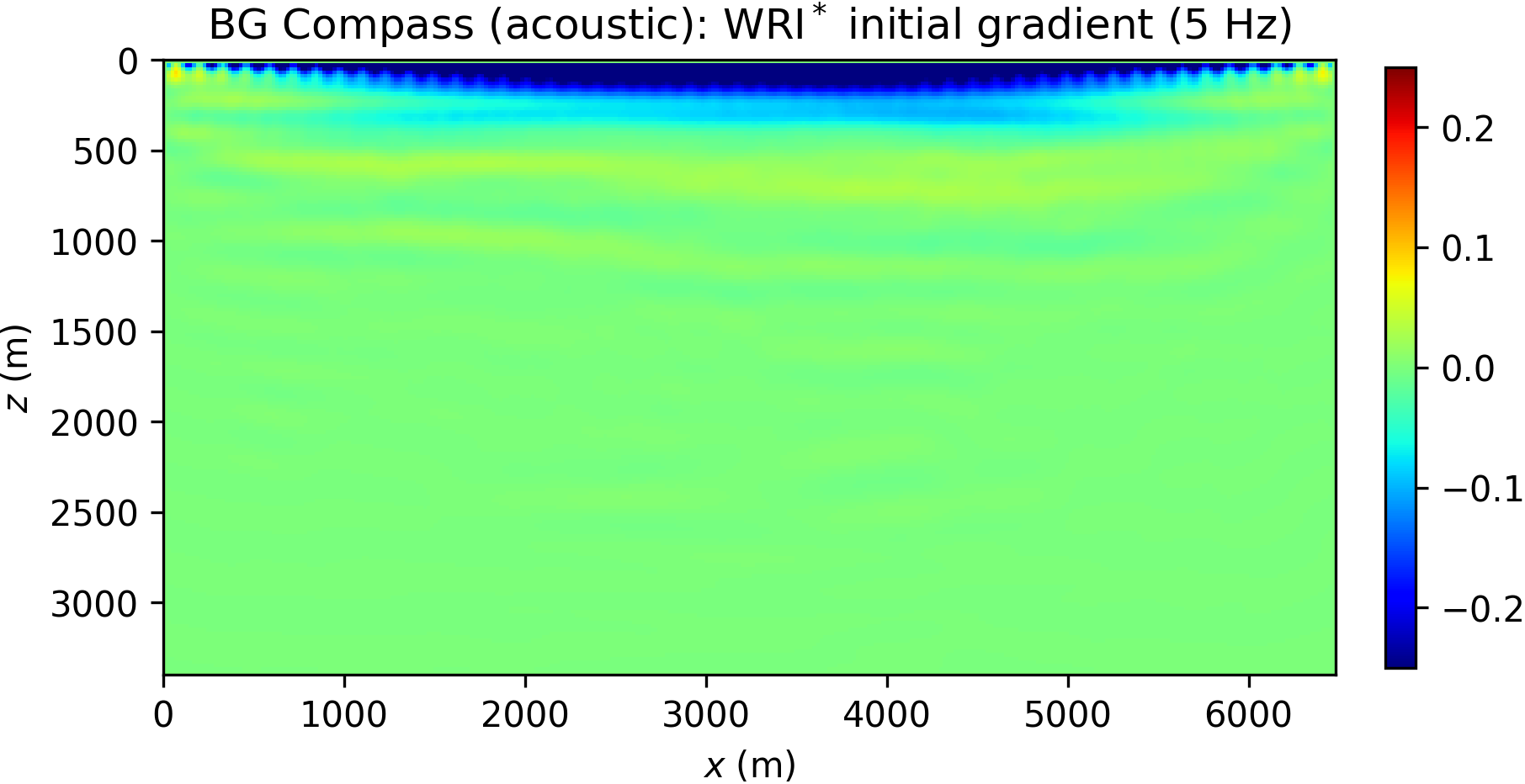}}
\caption{Gradient comparison for a single frequency component 5 Hz (with
respect to homogeneous squared slowness parameters): (a) FWI
(normalized), (b) WRI$\ast$ (normalized). The FWI gradient seeks to
update the water layer in the wrong direction.}\label{fig:BGCgradac}
\end{figure}

We model 51 sources with a Ricker wavelet (peak frequency 10 Hz), and
record data with dense receiver coverage (sampling rate according to the
grid spacing), without additional noise. Note that the frequency
components below 5 Hz in the data are filtered out and not used during
inversion. The difficulties associated with the BG Compass model are
hinted in Figure~\ref{fig:BGCgradac}, where the squared slowness
gradients of WRI$\ast$ and FWI (computed with respect to the background
model in Figure~\ref{fig:BGCmodelac-bg}) are compared. These are
obtained by filtering the data according to a narrow frequency band
centered at 5 Hz \citep[a similar result can be obtained by working
directly in the frequency domain,][]{Peters2014}. We observe that FWI
produces an erroneous update in the water layer, while WRI$\ast$ points
in the correct direction.

\begin{figure}
\centering
\subfloat[\label{fig:BGCinv-fwi}]{\includegraphics[width=1.000\hsize]{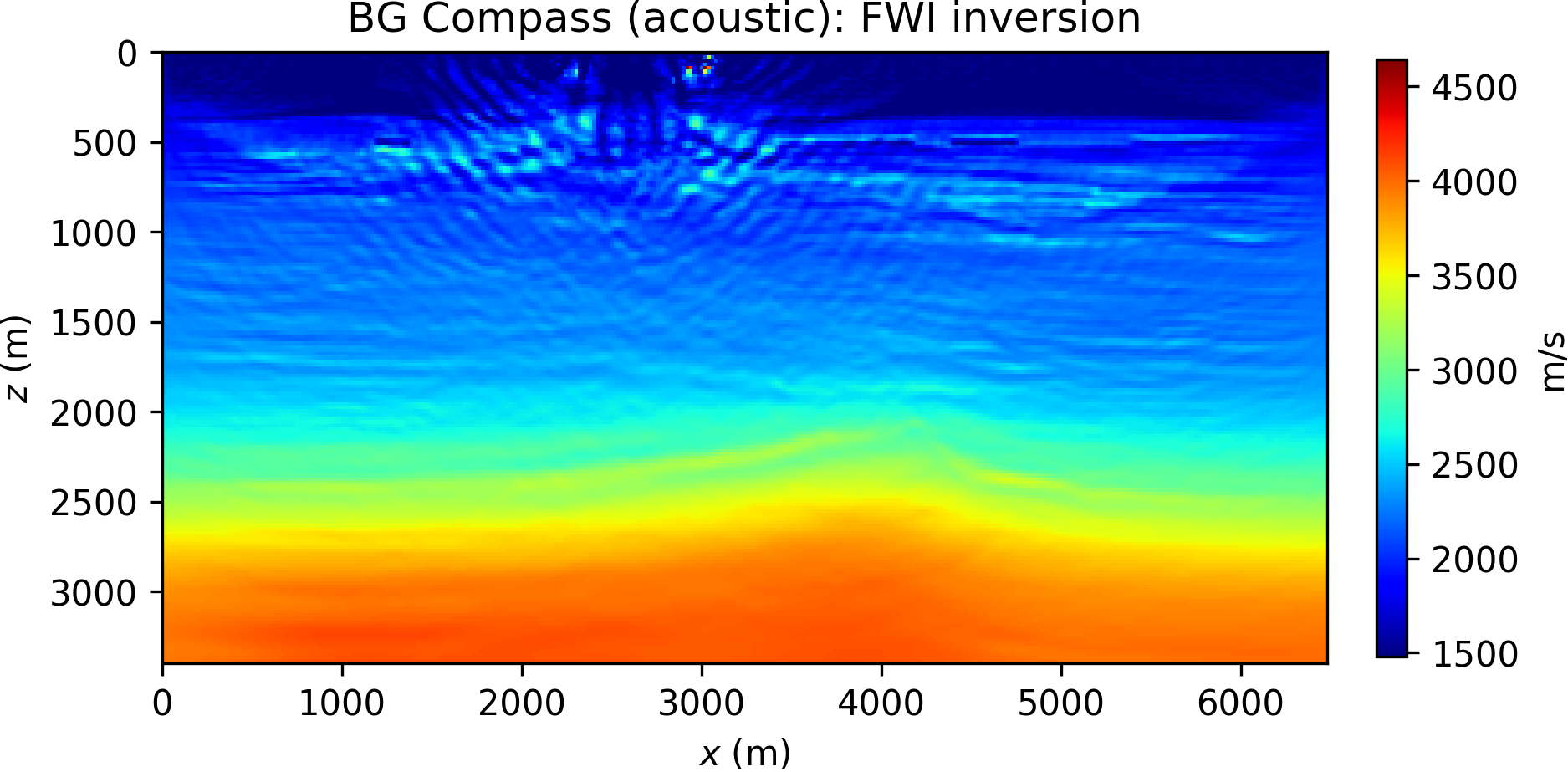}}
\\
\subfloat[\label{fig:BGCinv-twri}]{\includegraphics[width=1.000\hsize]{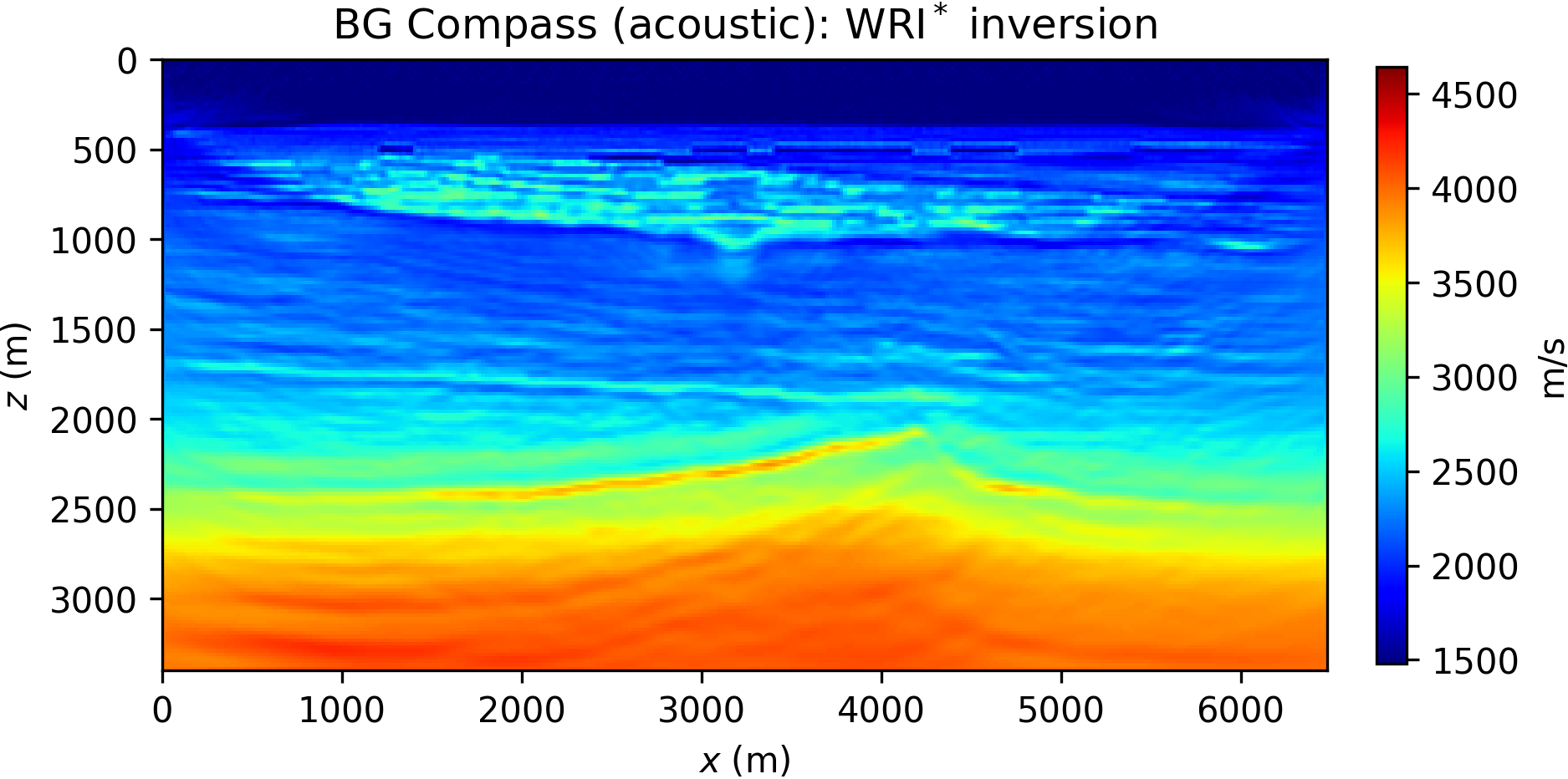}}
\caption{Acoustic BG Compass inversion result for: (a) FWI, (b)
WRI$\ast$.}\label{fig:BGCinv}
\end{figure}

For the inversion, we adopt a multiscale strategy \citep{bunksFWI}, by
filtering and inverting data components of increasing frequency content,
from 5 Hz to 20 Hz. This is achieved by preconditioning both data and
synthetics. For simplicity, we do not adapt the model grid to the
varying frequency. At each of these sequential stages, we perform l-BFGS
optimization for 20 iterations and use the result as the starting guess
for the next step. The process is repeated with multiple sweeps across
the frequency range, by repeating the multiscale inversion twice and
restarting from the lowest frequency when the first pass is completed.
Multiple sweeps are here necessary due
to the high-low velocity inversion specific of the BG Compass model.
During the first pass, the inversion develops a spurious reflector
across the high-velocity layer located between 500 m and 1000 m deep,
and subsequent passes are designed to remove
it. The final results for WRI$\ast$ and
FWI are collected in Figure~\ref{fig:BGCinv}. For reference, see also
\citet{Peters2014} for frequency-domain WRI inversion results on a
similar model.

Notably, the poor update generated by FWI within the water layer
eventually leads to the local minimum in Figure~\ref{fig:BGCinv-fwi}
(for FWI, we report the result after only one data sweep). The WRI$\ast$
result is shown in Figure~\ref{fig:BGCinv-twri}. Clearly, WRI$\ast$ is
able to correct for the water layer and resolve the
high-velocity/low-velocity sequence at 1 Km depth. The deeper part of
the model can be improved, in principle, by further optimization or
pseudo-Hessian preconditioning via incident field energy \citep[see][
for a simple Gauss-Newton scheme]{van2013mitigating}.

\begin{figure}
\centering
\subfloat[\label{fig:BGCrtm0}]{\includegraphics[width=1.000\hsize]{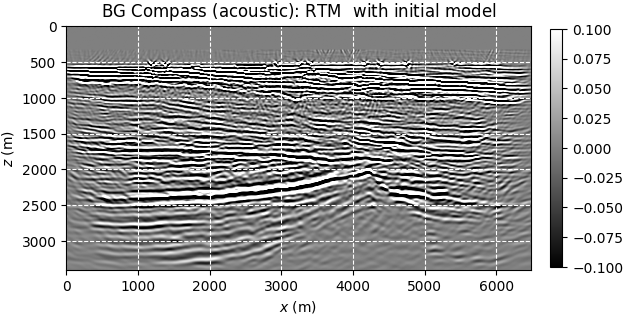}}
\\
\subfloat[\label{fig:BGCrtm-twri}]{\includegraphics[width=1.000\hsize]{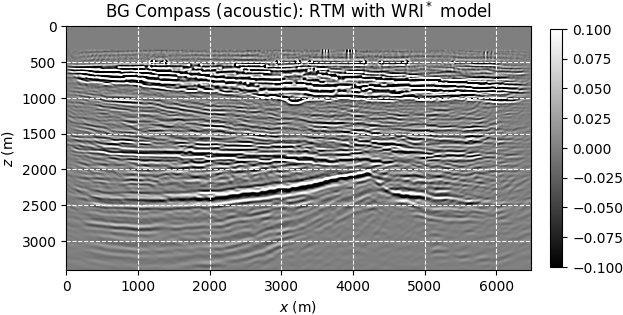}}
\caption{Reverse time migration results (normalized for display
purposes) for the BG Compass model, corresponding to different
background velocity models: (a) starting model (see
Figure~\ref{fig:BGCmodelac-bg}), (b) inversion result for WRI${\ast}$
(see Figure~\ref{fig:BGCinv-twri}). A grid is overlayed in white to aid the comparison.}\label{fig:BGCrtm}
\end{figure}

As a further quality control
assessment, in Figure~\ref{fig:BGCrtm} we compare RTM results obtained
from the initial velocity model and from the the WRI${\ast}$ inversion.
We migrate synthetic data generated with a Ricker wavelet of 20 Hz peak
frequency (while we used 10 Hz peak frequency data for inversion), the
rest of the source-receiver acquisition setting remains the same. The
RTM image obtained with the WRI${\ast}$ inverted model is resolved with
higher resolution at shallow depth, while the deeper reflectors more
accurately migrated.

\paragraph{Inversion with inaccurate modeling
assumptions}\label{inversion-with-inaccurate-modeling-assumptions}

\begin{figure}
\centering
\subfloat[\label{fig:BGCmodeltti-eps-true}]{\includegraphics[width=0.500\hsize]{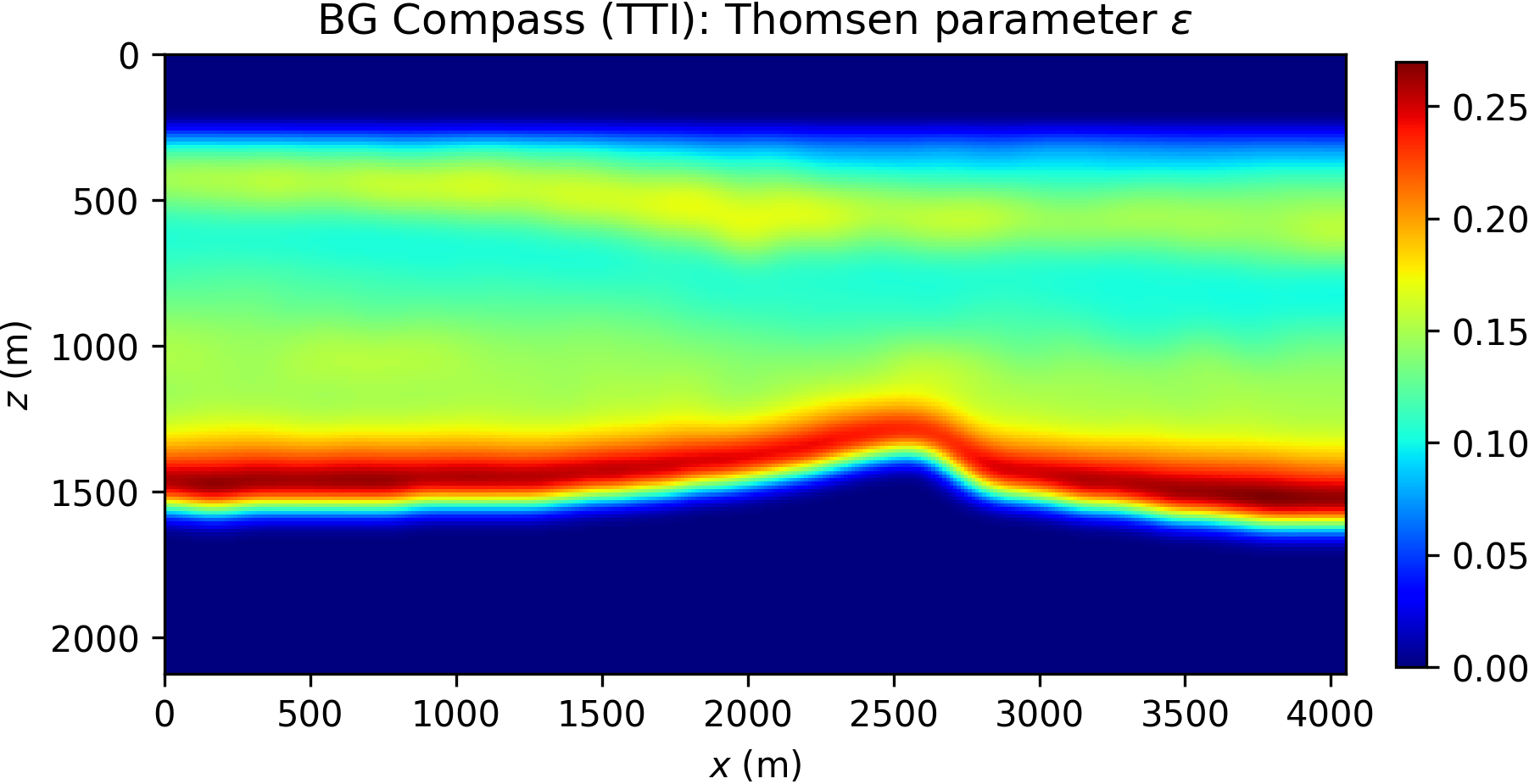}}
\\
\subfloat[\label{fig:BGCmodeltti-delta-true}]{\includegraphics[width=0.500\hsize]{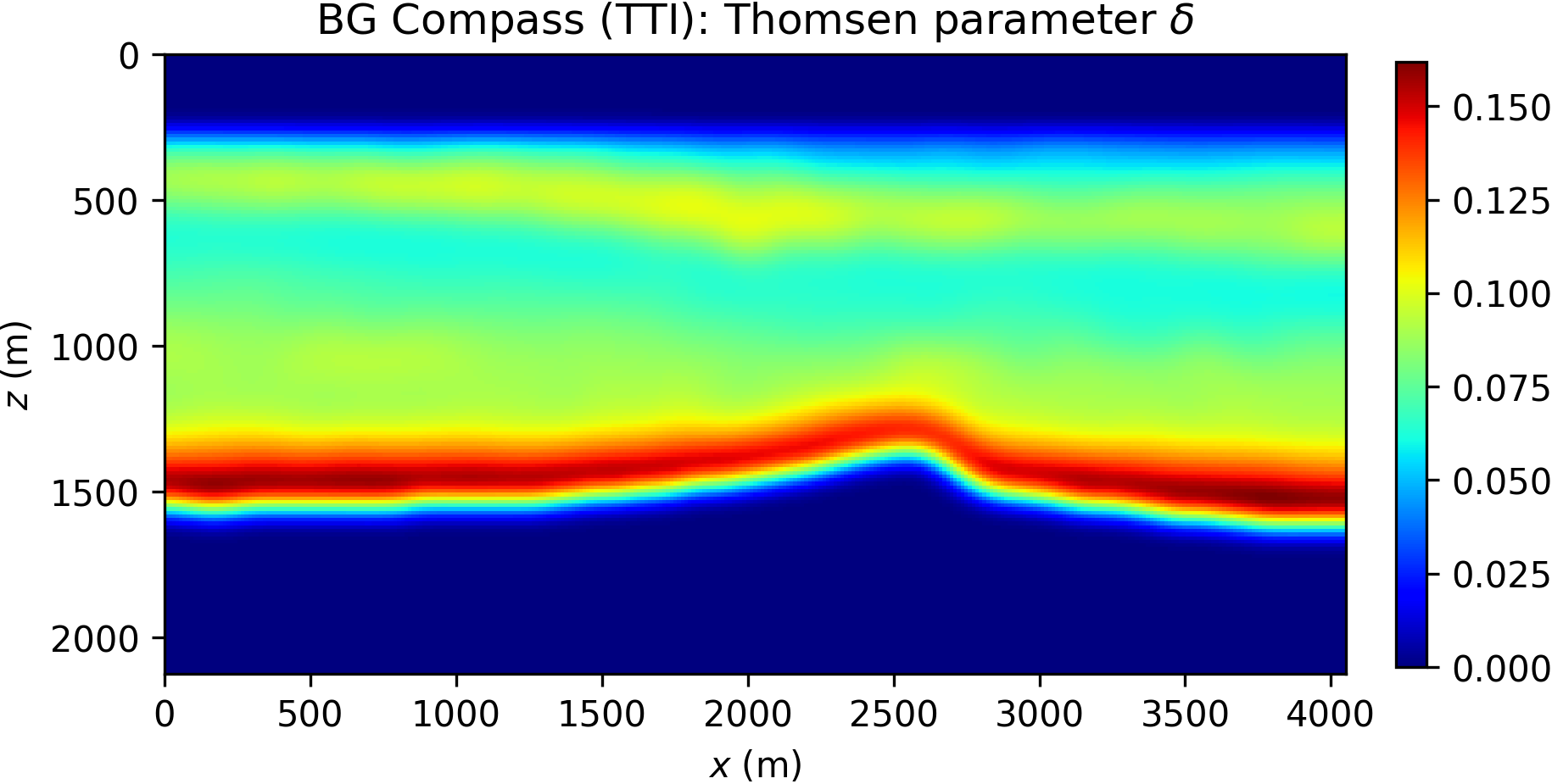}}
\\
\subfloat[\label{fig:BGCmodeltti-theta-true}]{\includegraphics[width=0.500\hsize]{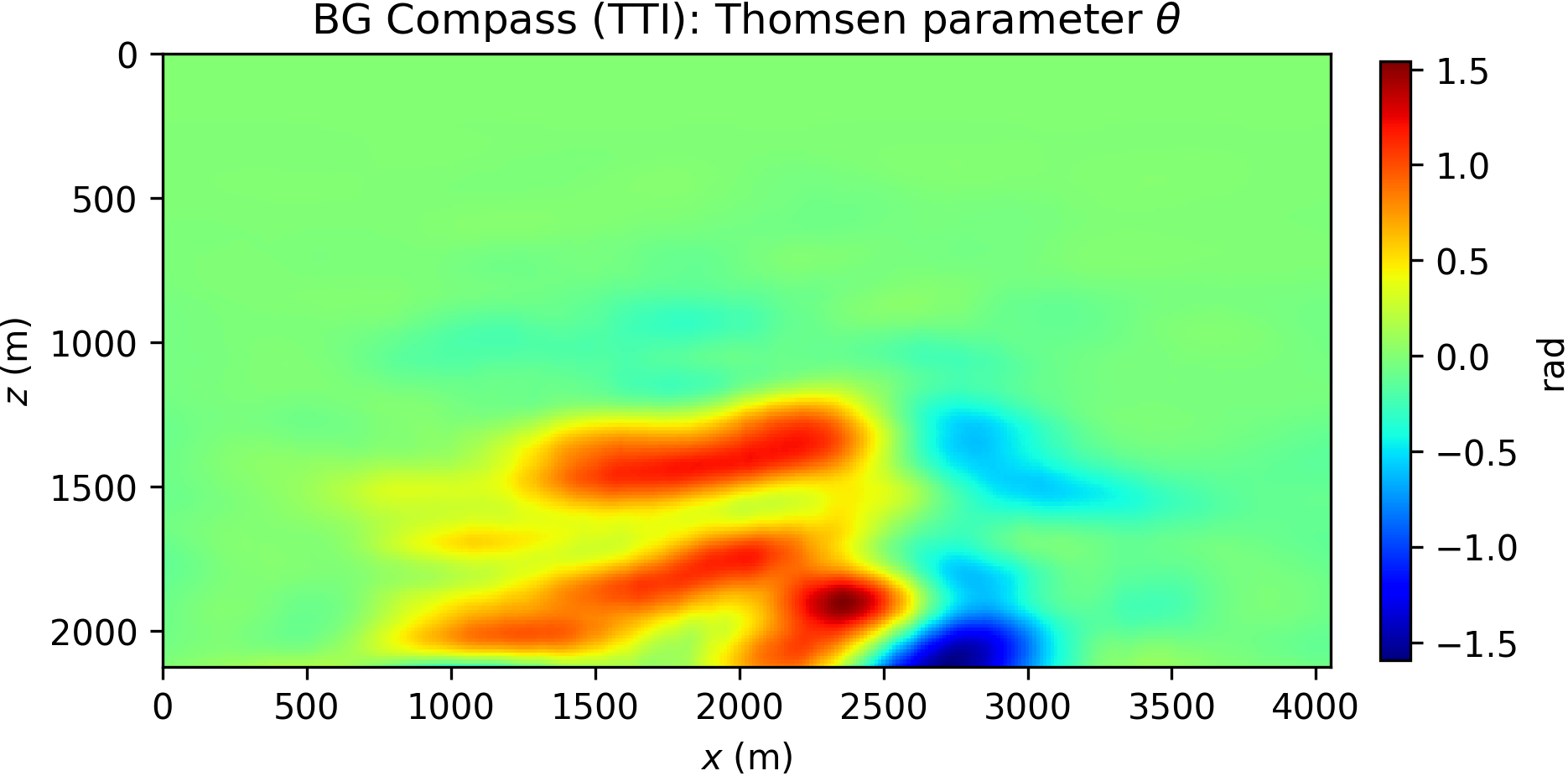}}
\caption{BG Compass, TTI model: (a) Thomsen parameter $\varepsilon$, (b)
Thomsen parameter $\delta$, (c) anisotropy dip angle $\theta$. Refer to
Figure~\ref{fig:BGCmodelac-true} for the velocity
model.}\label{fig:BGCmodeltti}
\end{figure}

Here, we consider a TTI anisotropic version of the acoustic BG Compass
model, previously shown in Figure~\ref{fig:BGCmodelac-true}. The model
is rescaled to simplify the problem by
mitigating the effect of the velocity kickback. Artifacts will now
appear deeper in the model. The Thomsen
parameters $\varepsilon$ and $\delta$ are synthesized from the velocity
model, and dip angles inferred from the orientation of the layers. The
TTI model is presented in Figure~\ref{fig:BGCmodeltti}.

Data are generated according to the same settings described for the
acoustic BG Compass model (note, however, that the TTI model size is
different).

\begin{figure}
\centering
\subfloat[\label{fig:BGCmodeltti-vel-bg}]{\includegraphics[width=0.500\hsize]{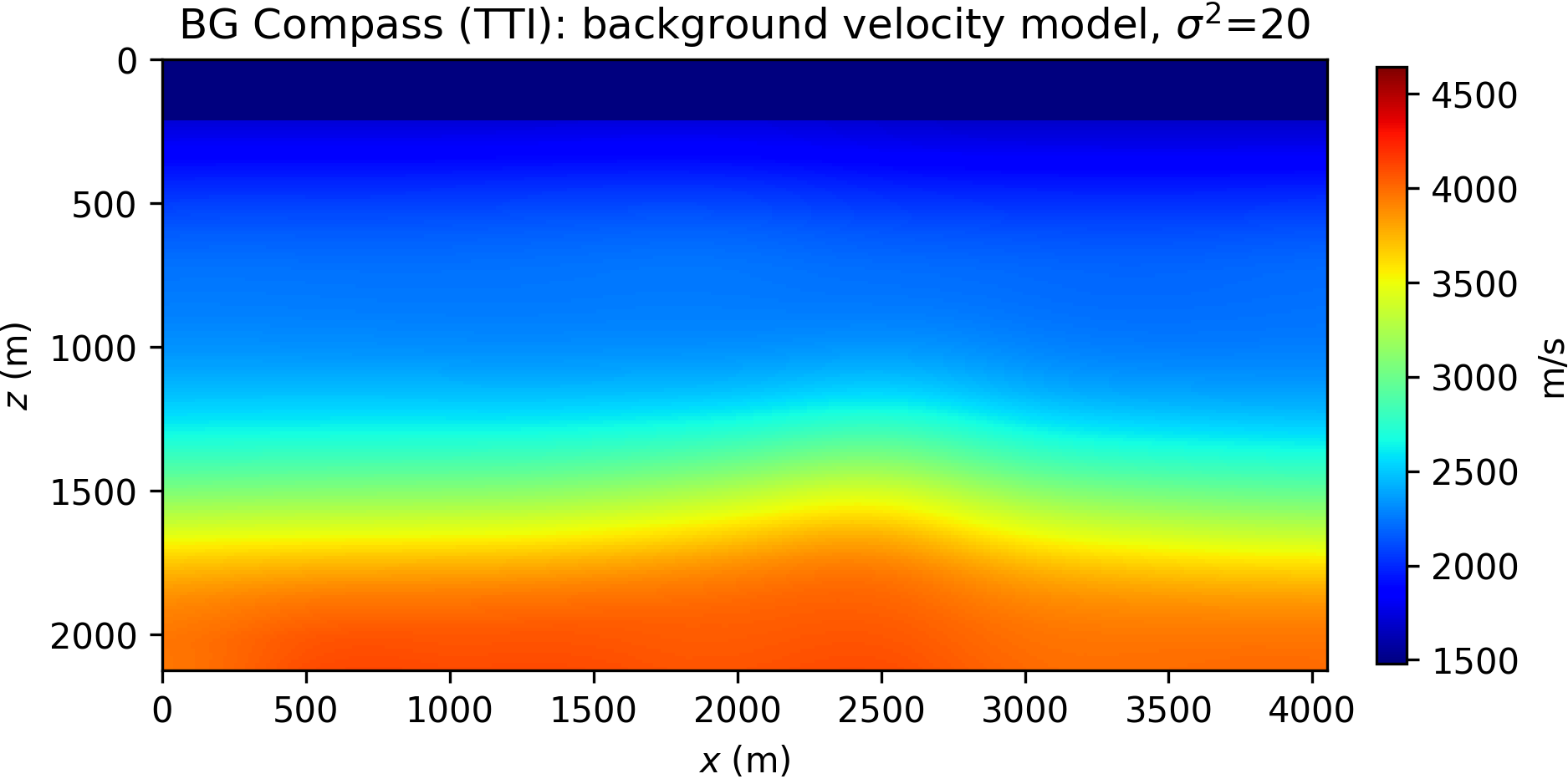}}
\subfloat[\label{fig:BGCmodeltti-eps-bg}]{\includegraphics[width=0.500\hsize]{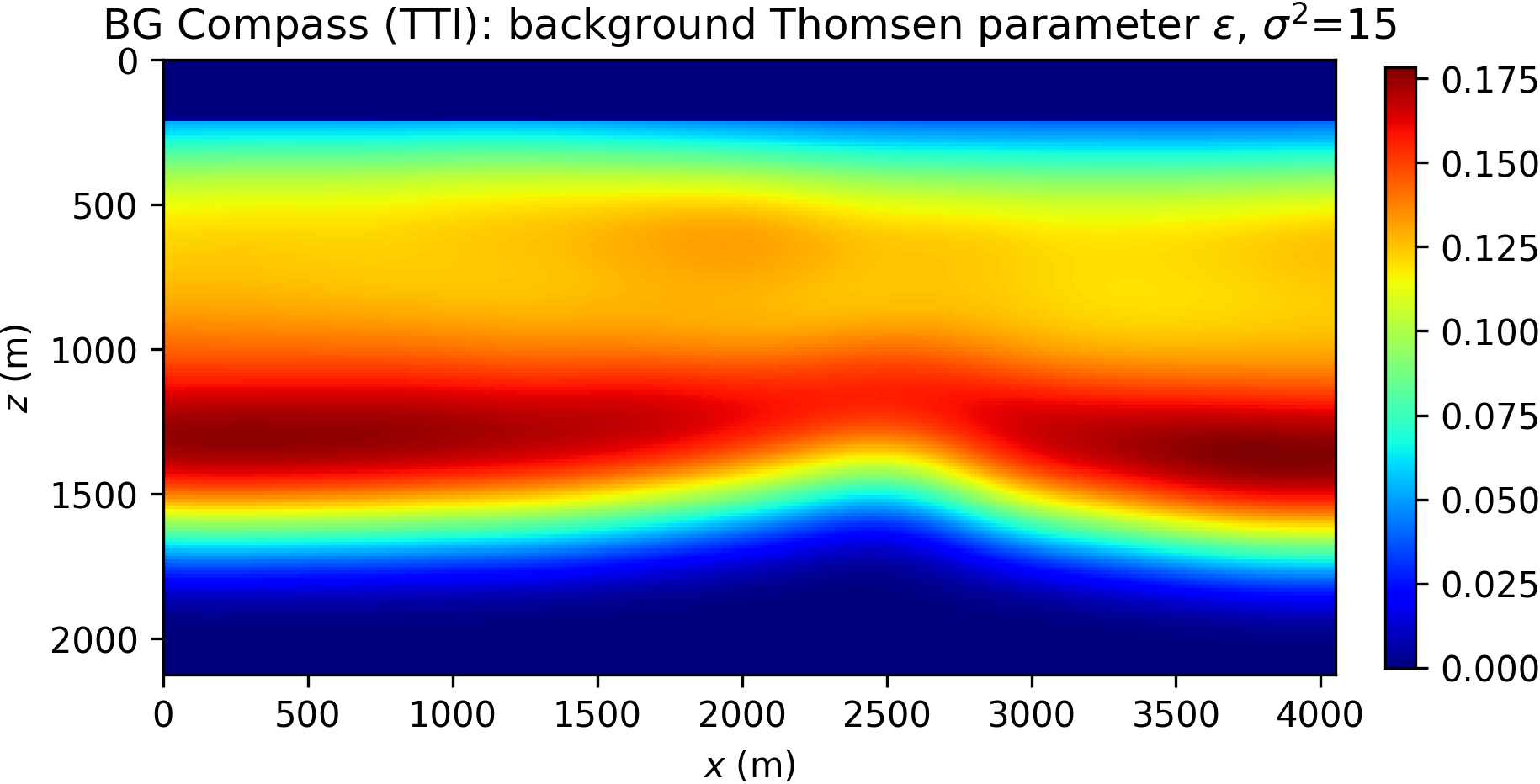}}
\\
\subfloat[\label{fig:BGCmodeltti-delta-bg}]{\includegraphics[width=0.500\hsize]{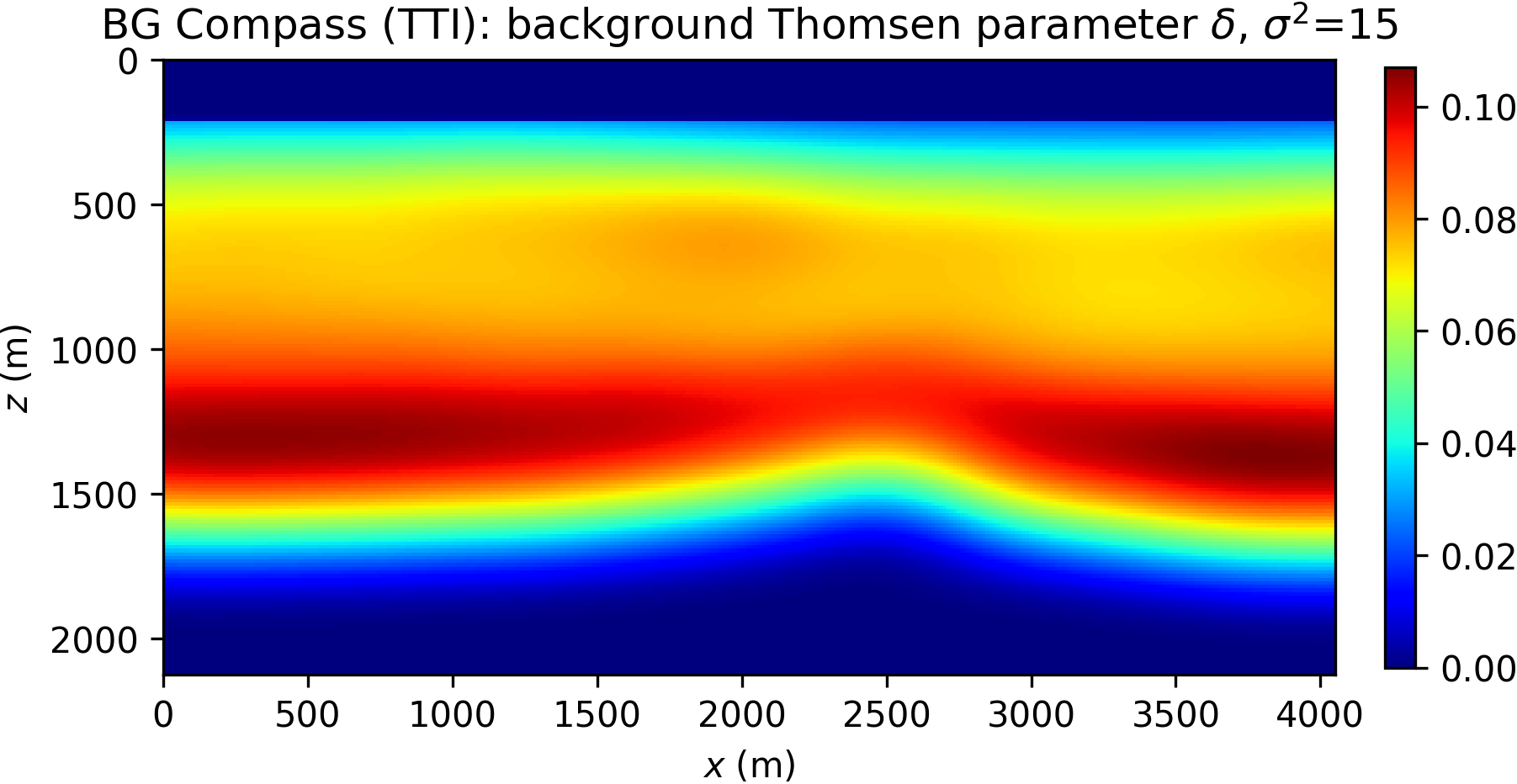}}
\subfloat[\label{fig:BGCmodeltti-theta-bg}]{\includegraphics[width=0.500\hsize]{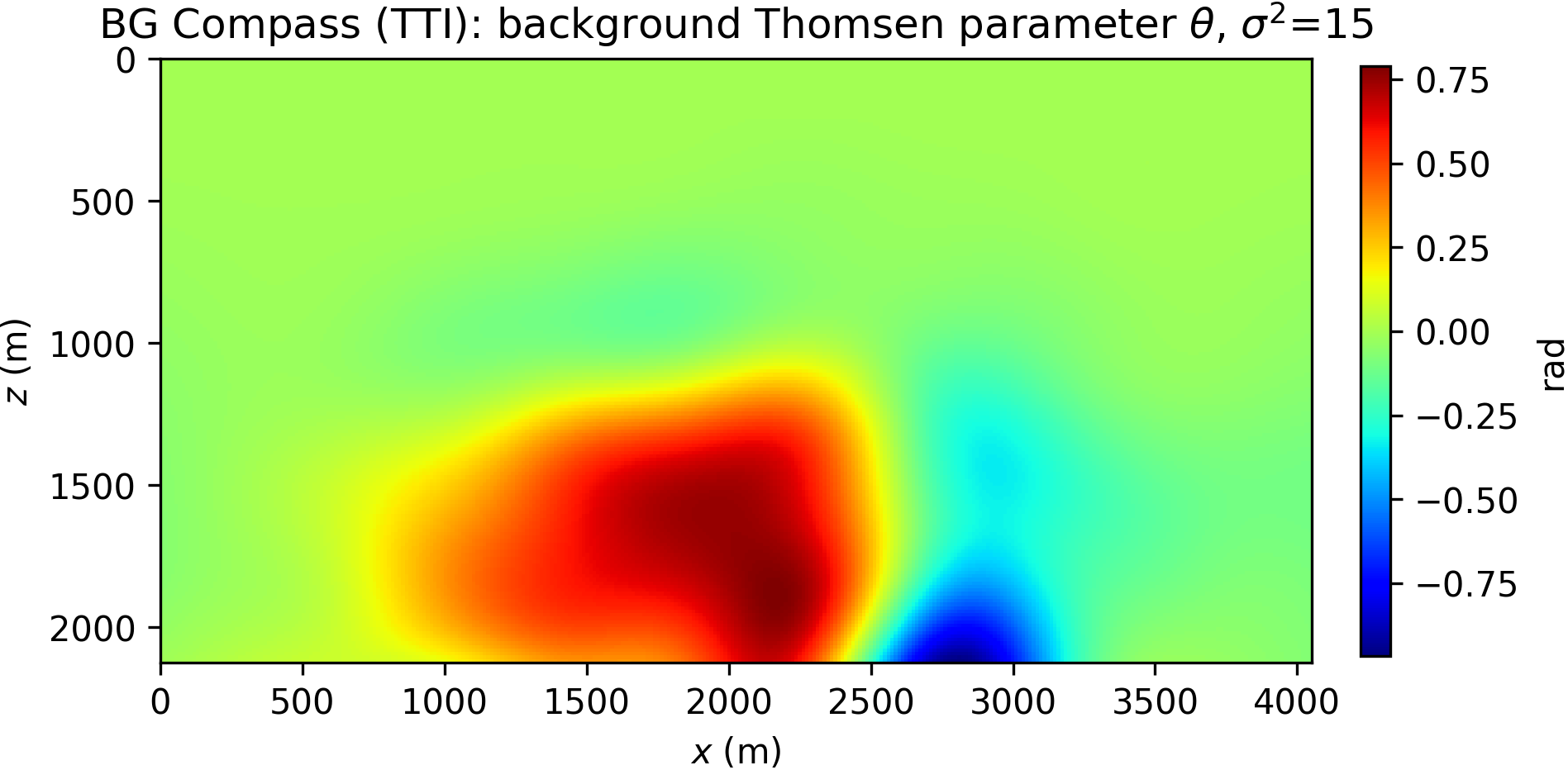}}
\caption{BG Compass, TTI starting model: (a) velocity model, (b) Thomsen
parameter $\varepsilon$, (c) Thomsen parameter $\delta$, (b) anisotropy
dip angle $\theta$. In the inversion, the anisotropic parameters
$\varepsilon$, $\delta$, and $\theta$ here displayed are kept fixed, and
only the velocity is updated. Note that the anisotropic parameters are
smoothed version of the true model in Figure~\ref{fig:BGCmodeltti}.
Smoothing is performed via Gaussian filtering with the indicated
standard deviation $\sigma$ (normalized with respect to the grid
spacing).}\label{fig:BGCmodelttibg}
\end{figure}

The inversion is performed with data generated with a Ricker wavelet of
peak frequency 10 Hz (with low frequencies filtered out), and the
optimization is carried out by 20 iterations of the spectral projected
gradient method (SPG) of \citet{birgin00} (the projection here bounding
velocity values). We compare WRI$\ast$ and FWI results by initializing
the inversion with smooth models
obtained from the true velocity map. The level of smoothness is
gradually increased to assess the robustness of the two
methods. The anisotropic parameters are
smoothed and kept fixed for all the experiments and throughout the
inversion. The inconsistency between true and assumed TTI parameters
will be the source of modeling inaccuracy with which we test the
resilience of the velocity model recovery with WRI$\ast$ and FWI.
Smoothing is carried out via Gaussian
filtering, modulated by the dimentionless standard deviation $\sigma$
obtained via normalization with the grid
spacing. An example of smoothed TTI model
is depicted in Figure~\ref{fig:BGCmodelttibg}.

\begin{figure}
\centering
\subfloat[\label{fig:BGCmodelttiinv-wrid-20}]{\includegraphics[width=0.500\hsize]{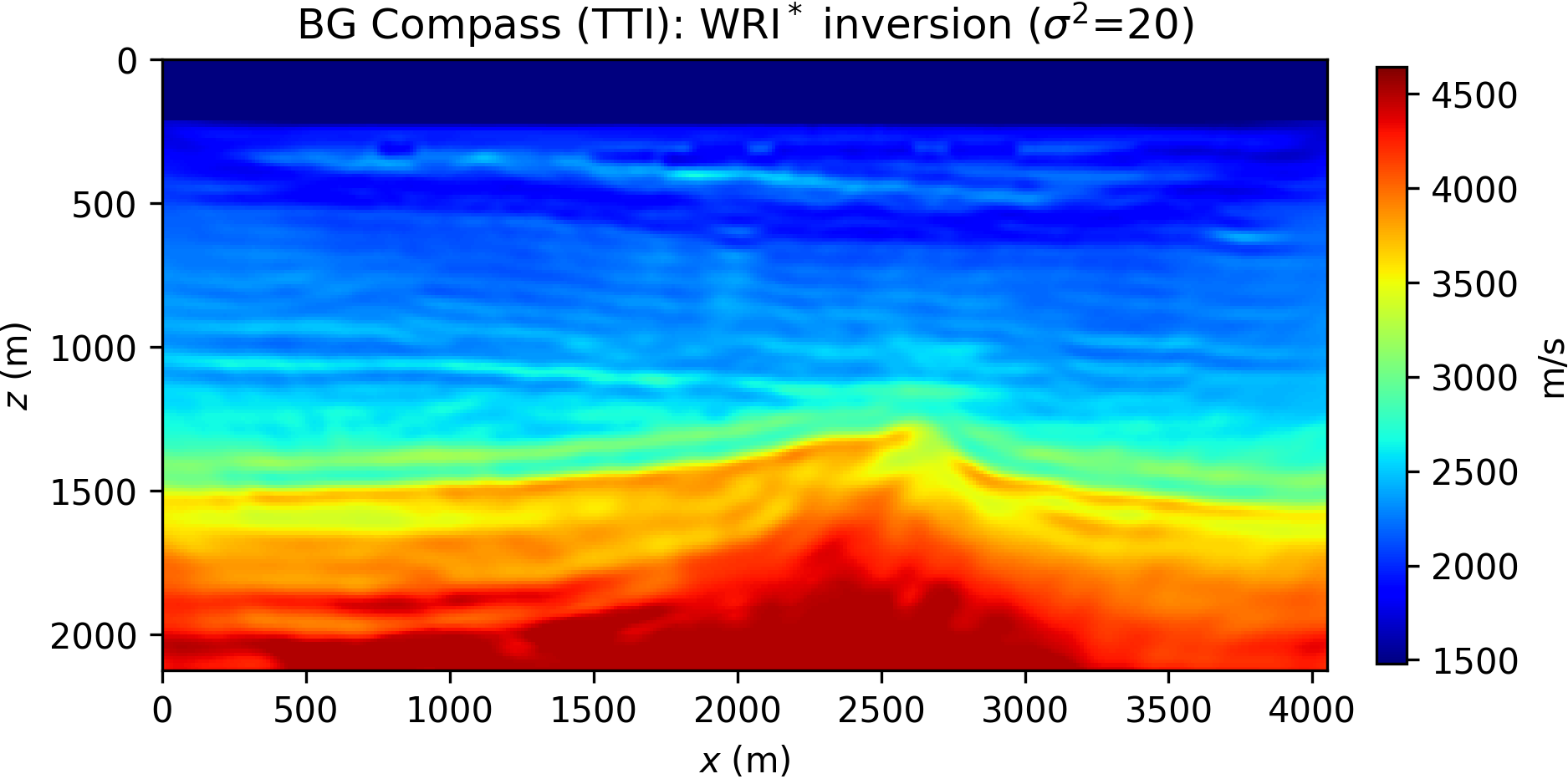}}
\subfloat[\label{fig:BGCmodelttiinv-fwi-20}]{\includegraphics[width=0.500\hsize]{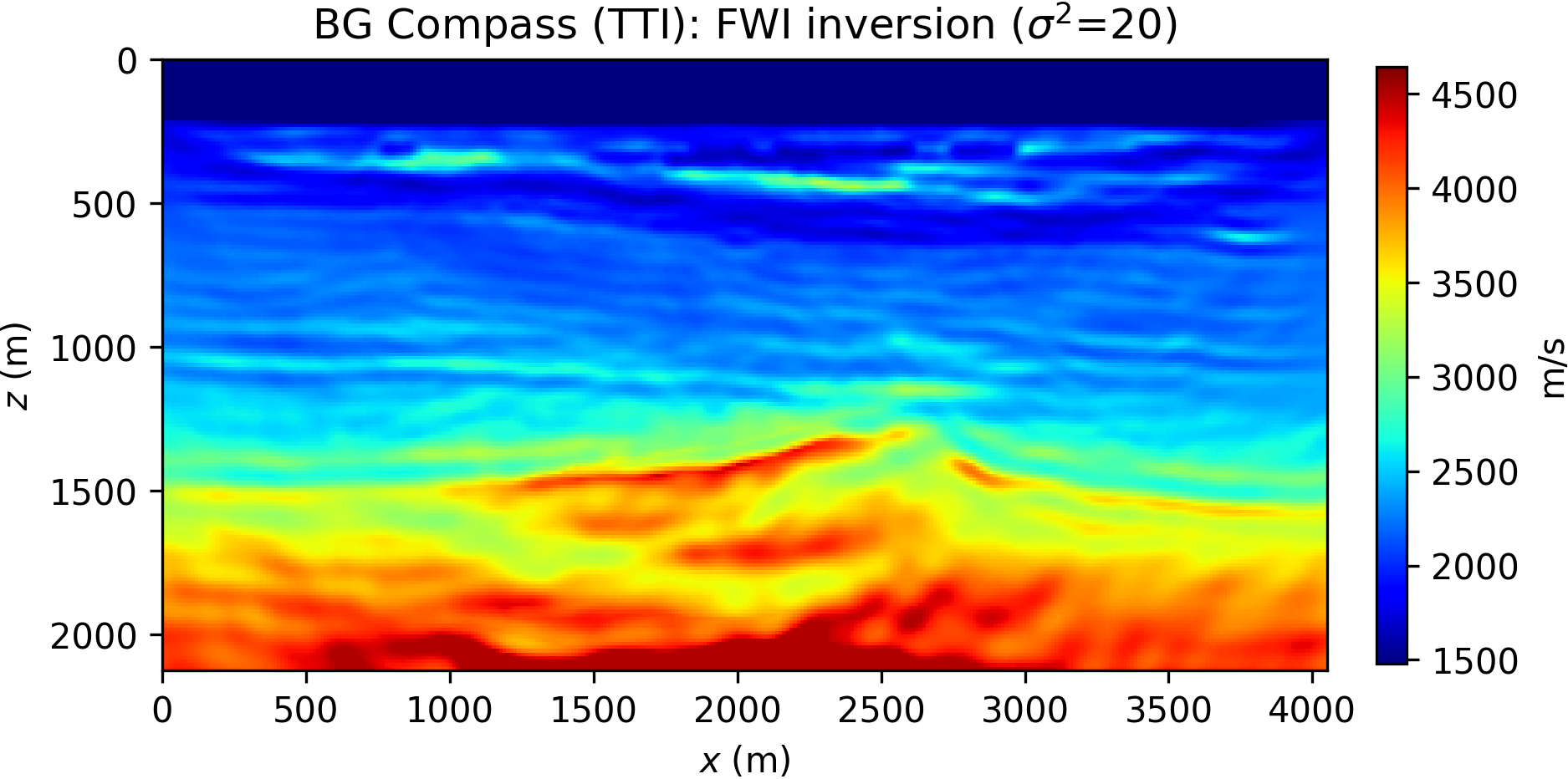}}
\\
\subfloat[\label{fig:BGCmodelttiinv-wrid-25}]{\includegraphics[width=0.500\hsize]{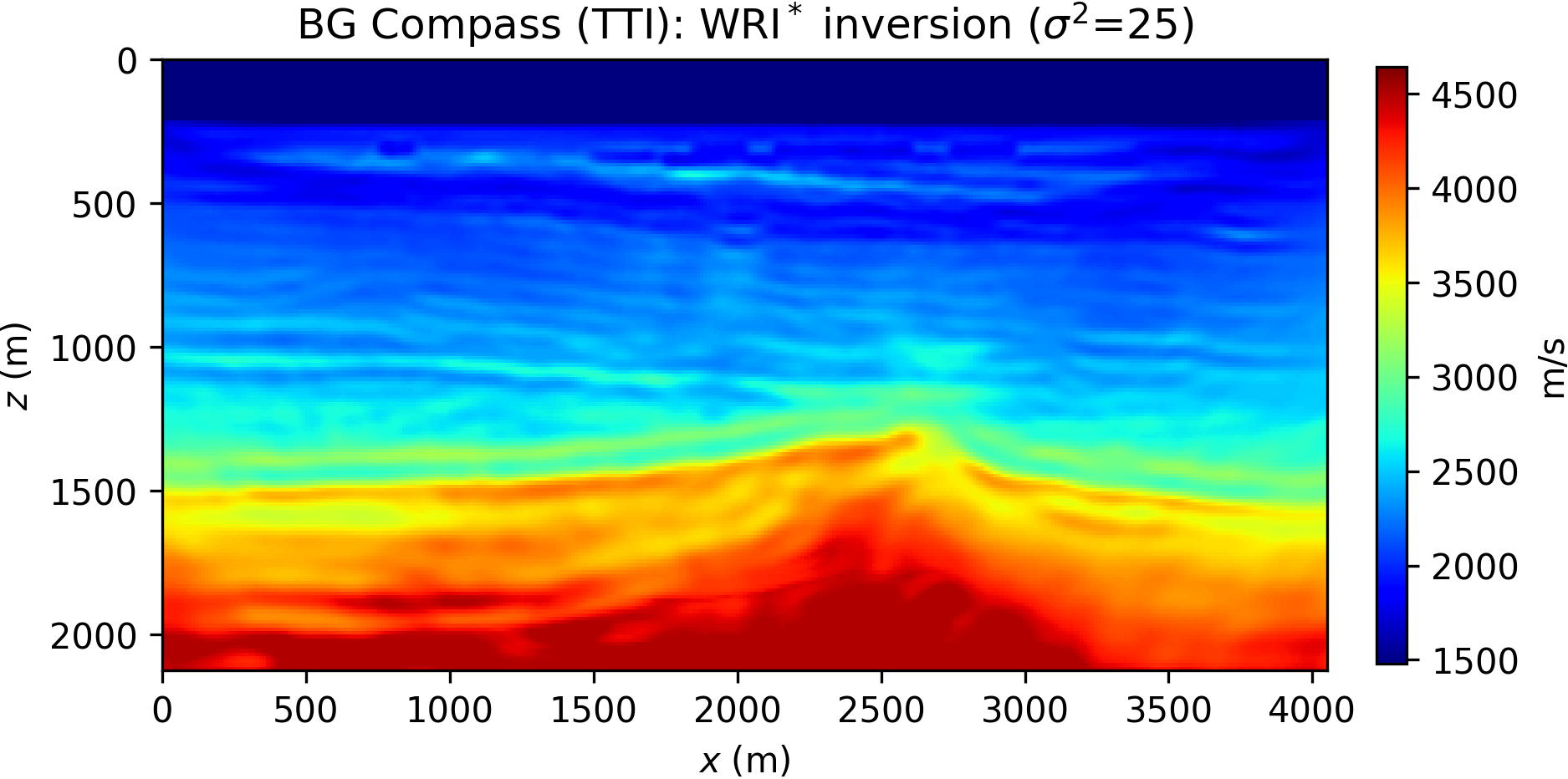}}
\subfloat[\label{fig:BGCmodelttiinv-fwi-25}]{\includegraphics[width=0.500\hsize]{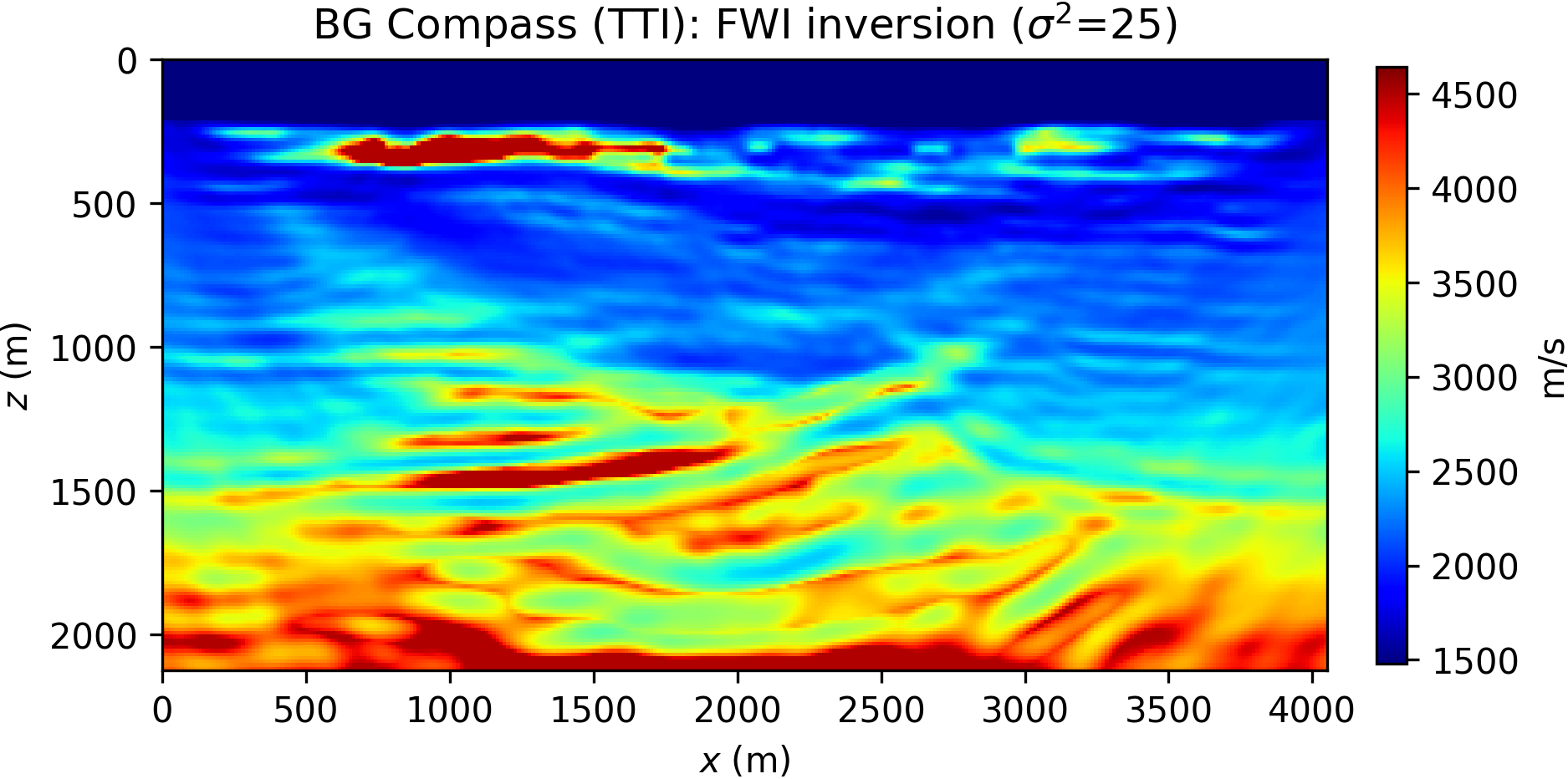}}
\\
\subfloat[\label{fig:BGCmodelttiinv-wrid-30}]{\includegraphics[width=0.500\hsize]{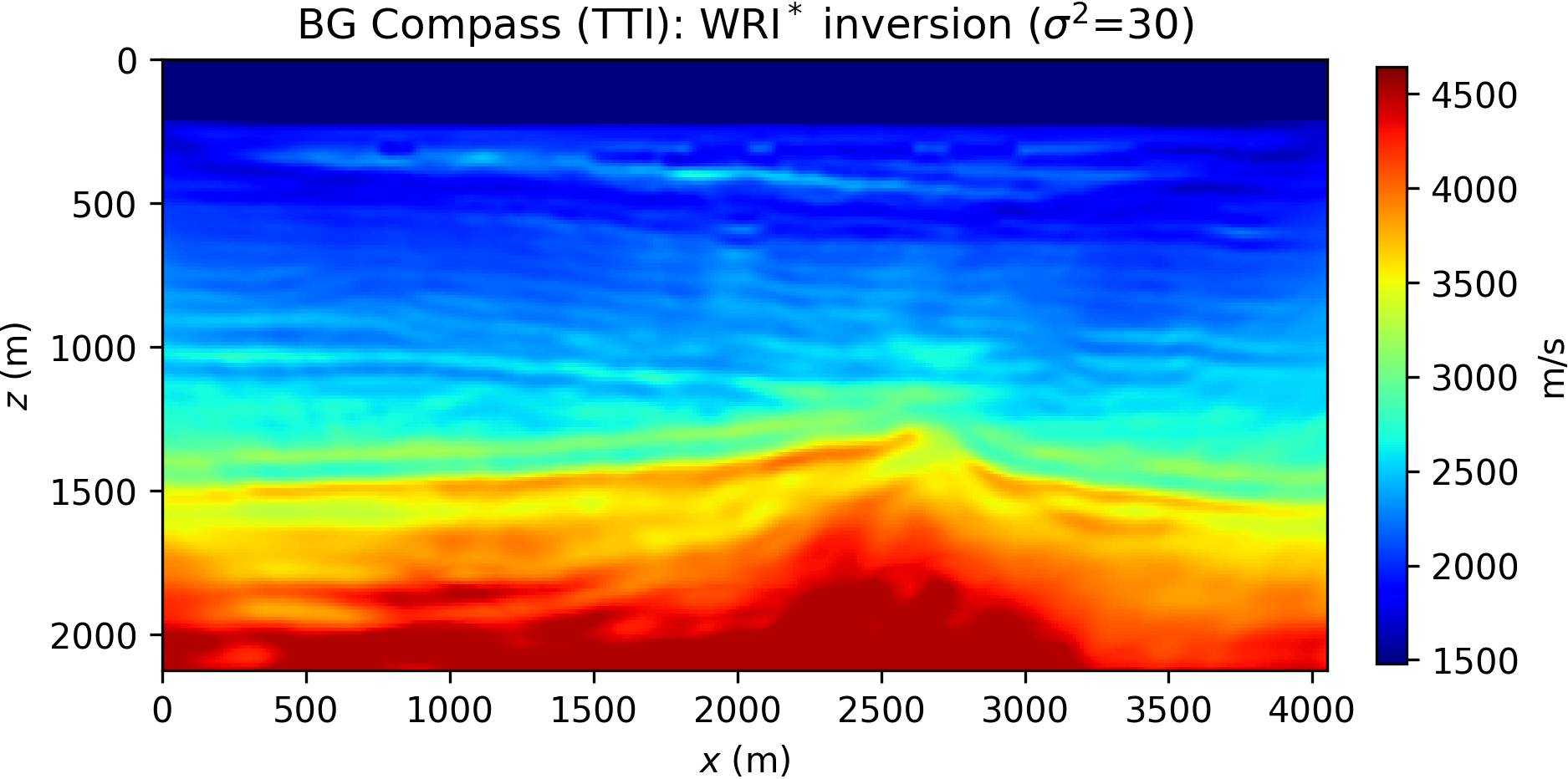}}
\subfloat[\label{fig:BGCmodelttiinv-fwi-30}]{\includegraphics[width=0.500\hsize]{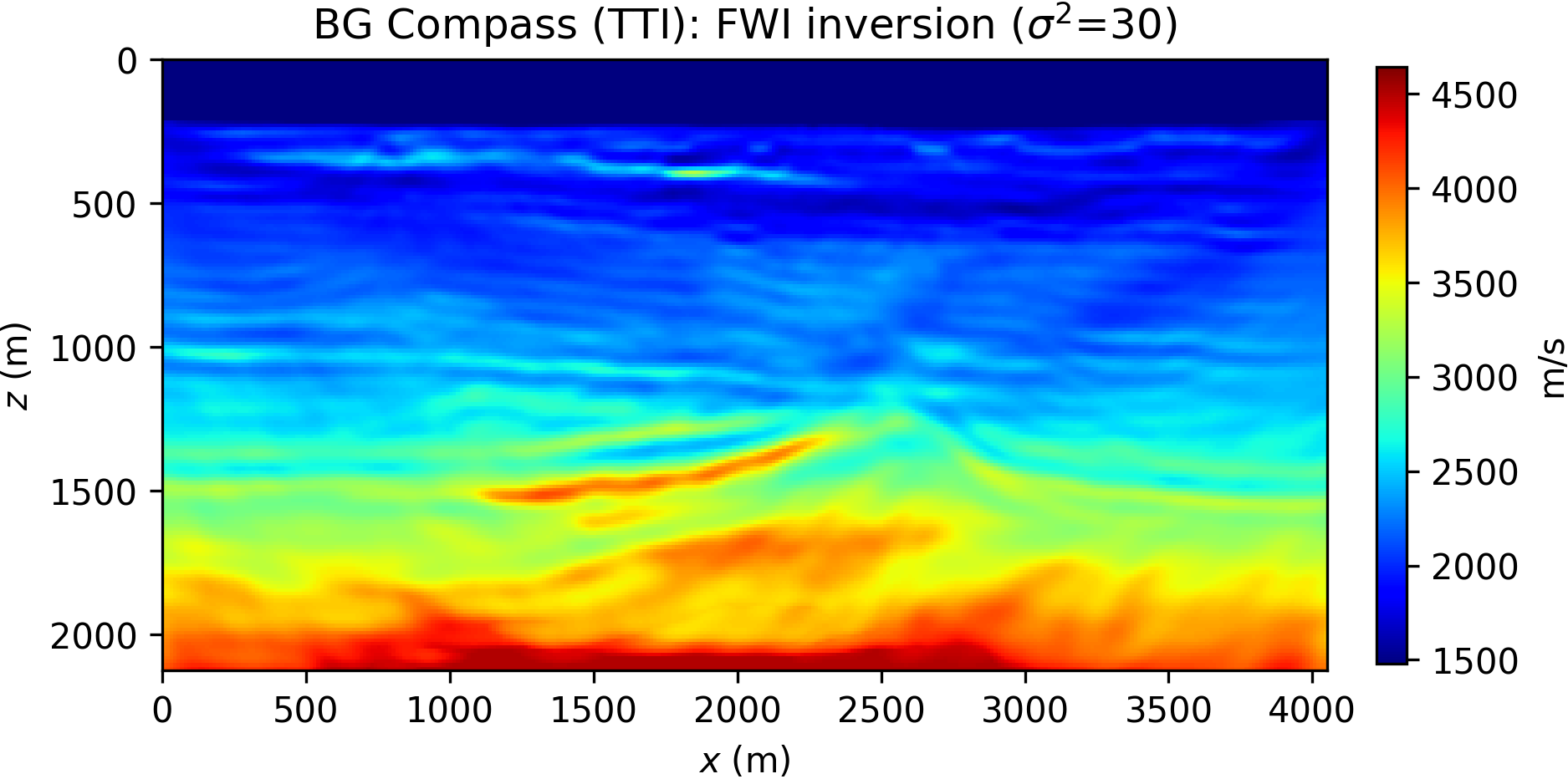}}
\\
\subfloat[\label{fig:BGCmodelttiinv-wrid-40}]{\includegraphics[width=0.500\hsize]{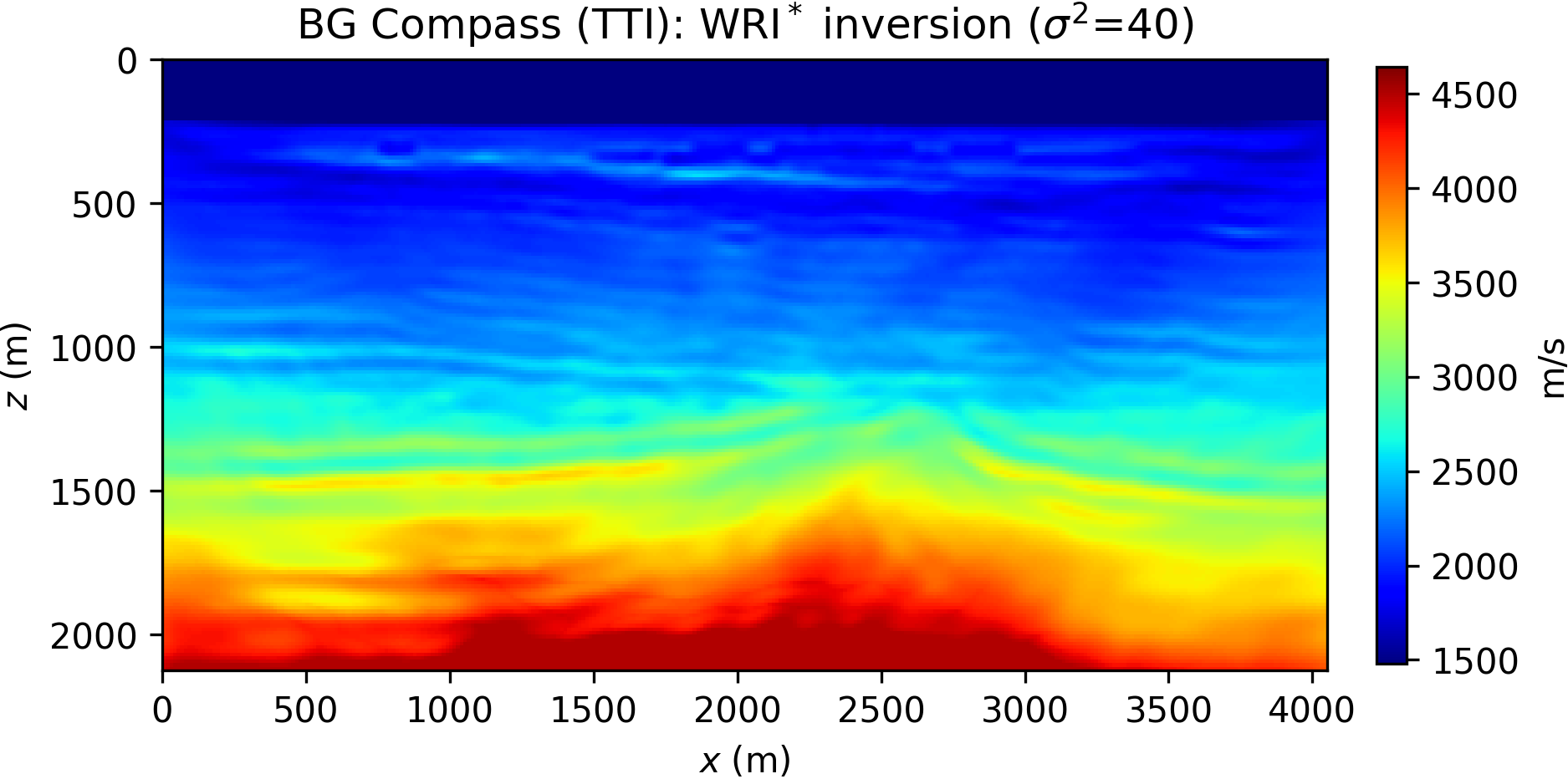}}
\subfloat[\label{fig:BGCmodelttiinv-fwi-40}]{\includegraphics[width=0.500\hsize]{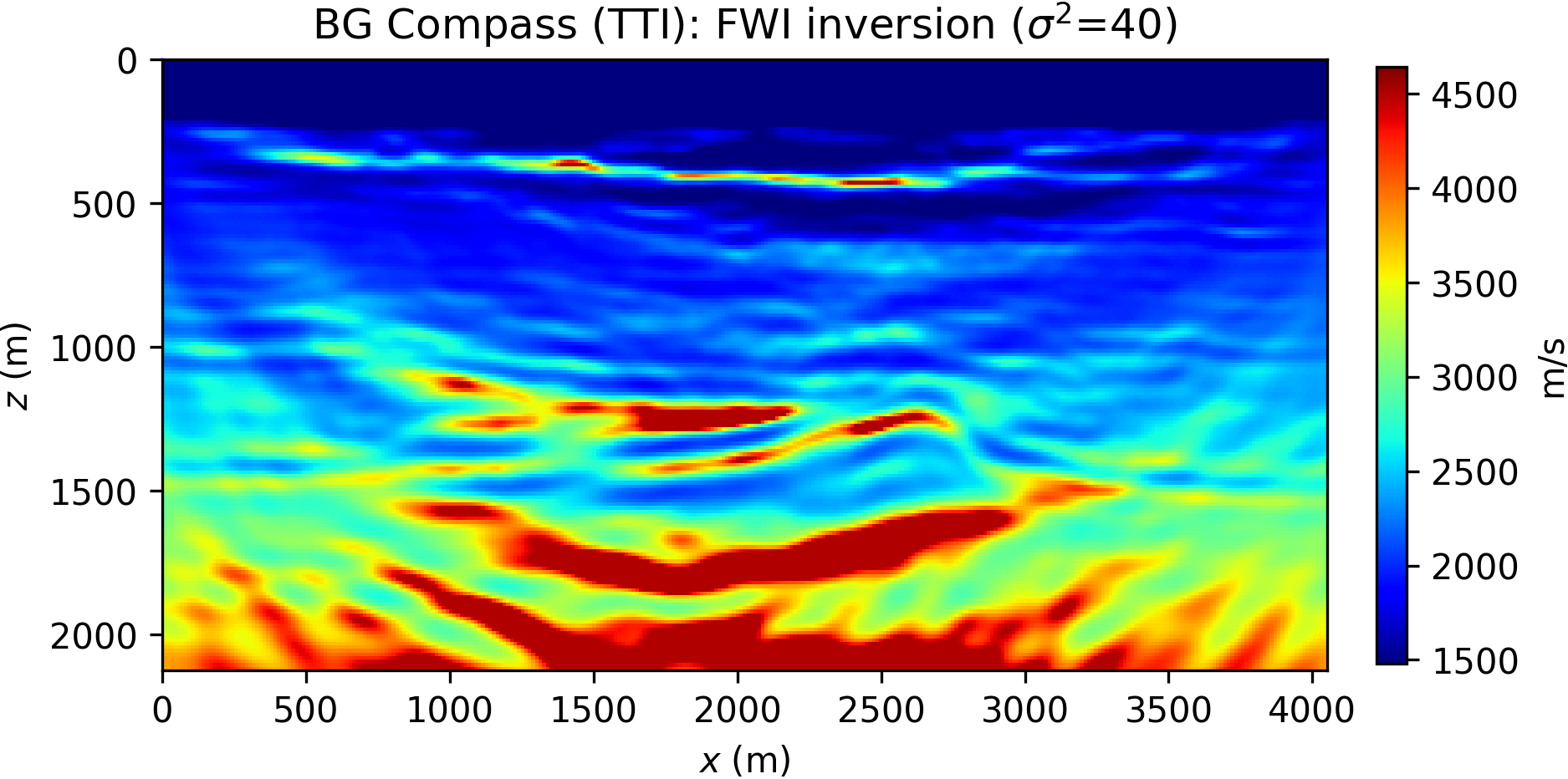}}
\caption{BG Compass, TTI inversion results acquired with starting
velocity models obtained from smoothing the ground truth with Gaussian
filters with varying standard deviations $\sigma$ normalized by grid
spacing (the anisotropy parameters are kept fixed as in
Figures~\ref{fig:BGCmodeltti-eps-bg}, \ref{fig:BGCmodeltti-delta-bg},
\ref{fig:BGCmodeltti-theta-bg}): (a,b) WRI$\ast$ and FWI results
(respectively) for $\sigma^2=20$, (c,d) $\sigma^2=25$, (e,f)
$\sigma^2=30$, (g,h) $\sigma^2=40$. The physical model underlying the
inversion always assume inaccurate anisotropy, but WRI$\ast$ delivers
consistently better result than FWI and remain relatively stable with
respect to the starting guess. FWI, on the other hand, does increasingly
get worse as the starting model is further from the ground
truth.}\label{fig:BGCmodelttiinv}
\end{figure}

The inversion results for WRI$\ast$ and FWI are displayed in
Figure~\ref{fig:BGCmodelttiinv}. We notice that the starting model and
wrong assumptions about anisotropy have an effect on WRI$\ast$ results,
especially in resolving the shallow high-velocity layer and the deeper
part of the model. Despite this, the WRI$\ast$ results remain relatively
stable with respect to the starting guess, the major difference being a
loss of resolution with depth with worse starting models. More notably,
they are consistently better than the FWI results, which are instead
heavily affected by this choice.

\subsection{Preliminary 3D results}\label{preliminary-3d-results}

\begin{figure}
\centering
\subfloat[\label{fig:lens3Dtrue}]{\includegraphics[width=0.500\hsize]{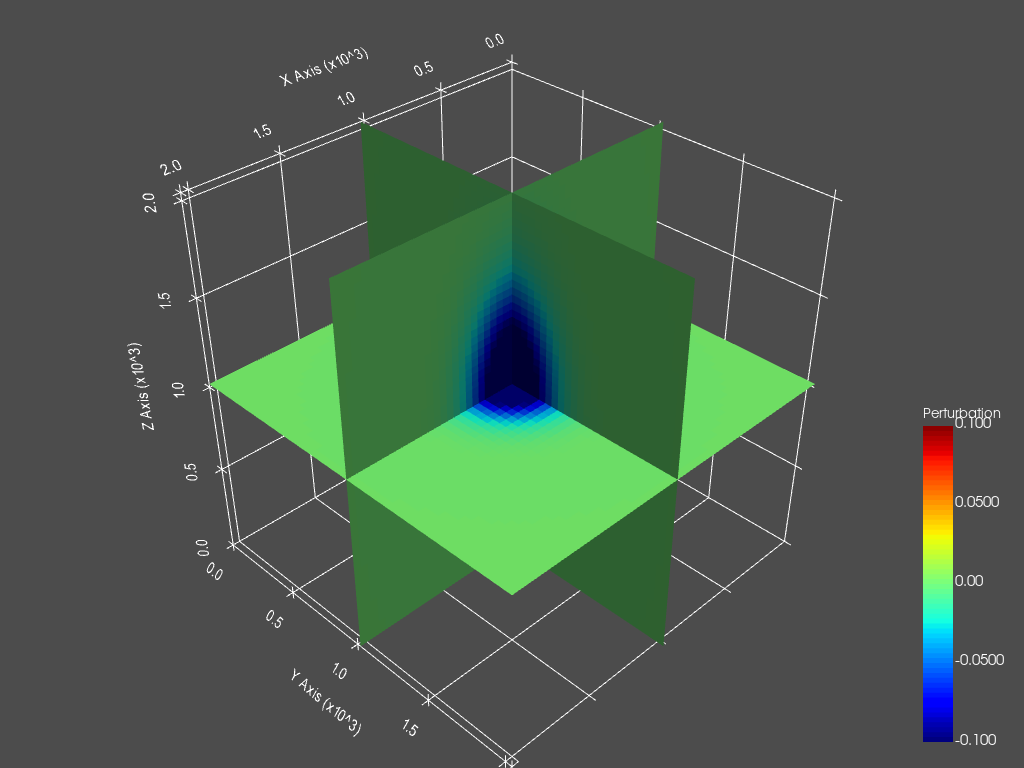}}
\\
\subfloat[\label{fig:lens3Dfwi}]{\includegraphics[width=0.500\hsize]{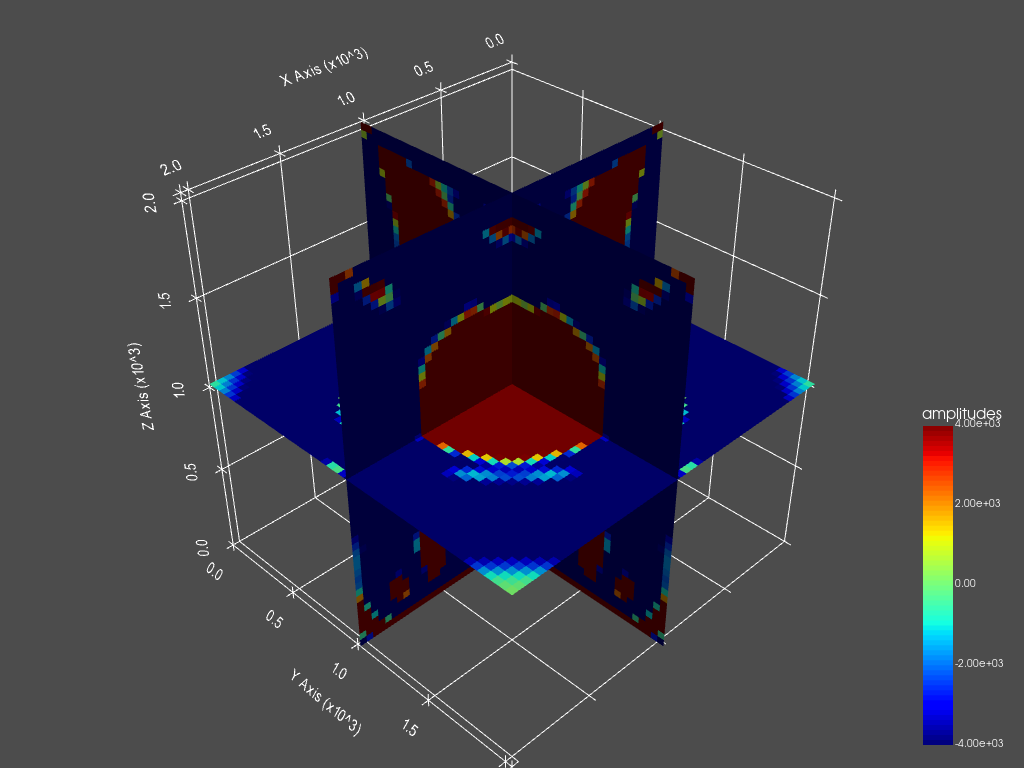}}
\\
\subfloat[\label{fig:lens3Dwri}]{\includegraphics[width=0.500\hsize]{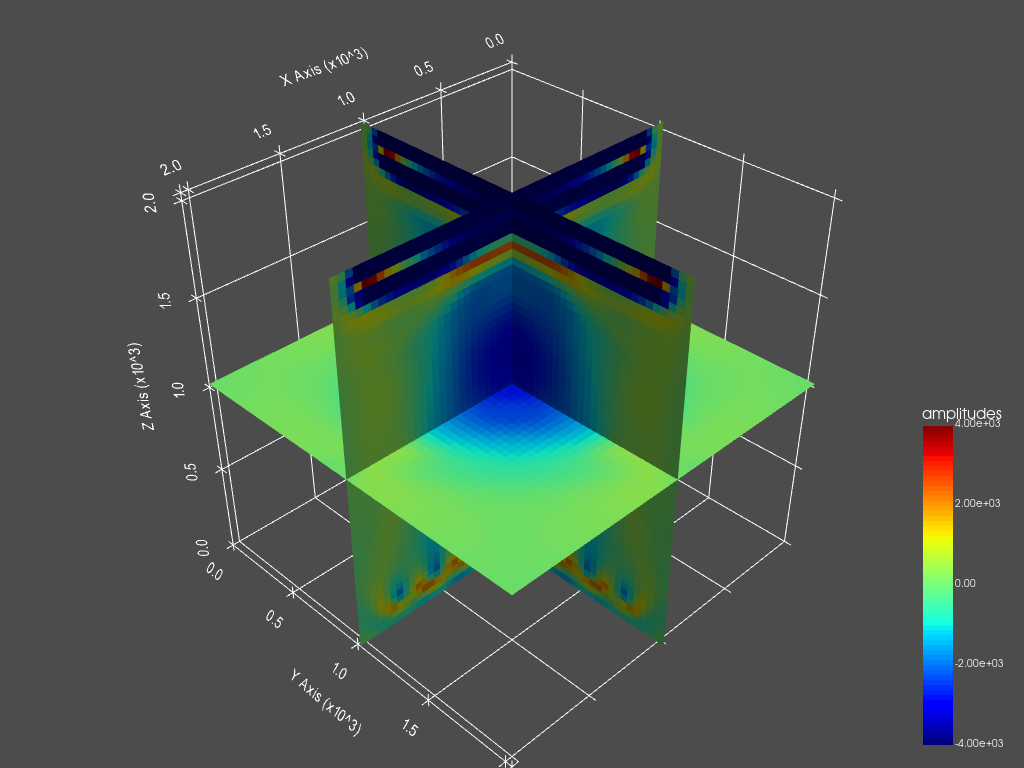}}
\caption{3D low-velocity lens model: (a) true model perturbation (with
respect to an homogeneous background), (b) initial gradient for FWI, (c)
initial gradient for WRI$\ast$. As in the 2D case, the FWI gradient
points to the wrong direction, while the WRI$\ast$ gradient allows for
the correct update.}\label{fig:lens3D}
\end{figure}

This numerical experiment aims at the gradient comparison between WRI$\ast$ and FWI for a small acoustic 3D
problem modeled after the previous low-velocity lens example in 2D, now
defined on a $51\times51\times51$ grid. We consider transmission data
generated by $8\times8$ evenly sampled sources located at $z=120$ m, and
$91\times91$ receivers at $z=1890$ m. We employ a Ricker source wavelet
of 5 Hz peak frequency, but the frequency components below 3 Hz are
filtered out. In Figure~\ref{fig:lens3Dtrue}, we depict the difference between the true model and an initial homogeneous model.

As in the 2D case, the gradient
comparison for WRI$\ast$ and FWI in Figure~\ref{fig:lens3D} shows that
WRI$\ast$, unlike FWI, leads to a correct velocity model
update.

\section{Discussion}\label{discussion}

The application of modern seismic inversion tools to field data has been
curbed by the need for tremendous computational resources, which are
required for wave simulations on extremely large 3D grids. Moreover,
classical acoustic models are being discarded in favor of
pseudo-acoustic (now standard in
industrial imaging) or elastic simulators
to account for anisotropic phenomena in field data. Less simplistic
modeling inevitably adds to the computational complexity of the problem.
Be that as it may, the oil and gas industry manages to employ FWI on a
routine basis, thanks to efficient time-domain simulators, and its
business value is widely acknowledged. Unfortunately, the success of FWI
is highly dependent on a lengthy preprocessing phase aimed at the
retrieval of a kinematically accurate background and starting guess for
seismic inversion.

The main objective of this work is to dispense with the need for
accurate initial velocity models in seismic inversion. Despite the fact
that many solutions exist focusing on robustness with respect to local
minima, they do not scale efficiently with grid size and are not
relevant for real-world applications. One such example is WRI. WRI is,
nonetheless, an attractive method that is not only robust with respect
to the starting guess, but also has the potential to handle moderate
inaccuracies in some of the physical parameters that are typically kept
fixed during inversion. Indeed, WRI hinges on a relaxed wave equation,
whose solution, however, can only be approximated iteratively in 3D with
considerable computational costs.

We presented a novel method, WRI$\ast$ which is both scalable and
robust. It is based on a denoising reformulation of WRI, which amounts
to the minimization of the wave equation error subject to the data
misfit being less than a known noise level. The resulting saddle-point
problem involves the usual model unknowns, such as squared slowness, and
Lagrangian multipliers having the same dimension as data. We stress that
all the computations needed for the evaluation of the Lagrangian and its
gradients require standard modeling in time domain, which can be
attained for realistically sized problems (contrary to frequency domain
formulations). Since the dual variables can be stored in memory, we
might in principle devise a joint optimization scheme. In this paper,
however, we approximated the Lagrangian multipliers by the scaled data
residual in order to reduce complexity (a more accurate assessment of
the trade-off between multiplier approximation and reconstruction
quality is left for future studies).

We demonstrated, through extensive numerical experimentation, that the
resulting method is consistently more robust than FWI against local
minima. We also tested the tolerance of the method toward faulty
modeling assumptions, e.g.~when seismic data obtained with anisotropic
modeling is being matched with predictions based on the wrong
anisotropic model. WRI$\ast$, which employs a relaxation of the wave
equation, produces better results than FWI, under the same conditions.

In general, the results obtained from WRI$\ast$ do depend on the
particular choice of hyperparameters such as the data noise level
$\epsilon$, defined in equation~\eqref{eq:denoise}. This parameter
governs the trade-off between wave equation error and data misfit and is
akin to the role of the weighting parameter $\lambda$, in
equation~\eqref{eq:WRI}, for conventional WRI. We find that decreasing
its value tends to produce lower resolution results. On the other hand,
choosing a small value for $\epsilon$ (or even $\epsilon=0$) results in
relaxed physics and is beneficial for local minimum avoidance. In our
experience, WRI$\ast$ retains the same ability of WRI to circumvent
local minima, but might produce less qualitative results, especially in
terms of resolution. We note that conventional WRI is not as affected as
WRI$\ast$ by the choice of $\lambda$. The fundamental reason for this
phenomenon is rooted in the inexact, but relatively inexpensive, dual
variable approximation in equation~\eqref{eq:alpha}, as opposed to
equation~\eqref{eq:WRIdual_yopt}. Taking into account the dependency of
the approximated dual variable with respect to the model parameter, as
in equation~\eqref{eq:WRIdg_gradm_corr}, is designed to partially
compensate for this behavior \citep{ablin2020superefficiency}. We then
advocate for either a continuation strategy for $\epsilon$, by
starting with $\epsilon=0$ and
increasing its value during the inversion (as hinted by the Marmousi
experiment in this paper), proper
preconditioning \citep[as in the Gauss-Newton approach
in][]{van2013mitigating}, or a sequential approach consisting of a
WRI$\ast$ phase followed by FWI 
(albeit, at the cost of potentially losing the robustness of WRI$\ast$
towards modeling hinging on inaccurate physical
parameters). Another approach might
consist of iteratively solving for the optimal dual variable, but we
envisage this can be only feasible in combination with stochastic
optimization, where simultaneous sources are considered.

From a computational perspective, WRI$\ast$ requires roughly twice as
much the cost needed for FWI, per objective evaluation or gradient
calculation. Given the gain in robustness previously discussed, we deem
this extra cost acceptable in the many situations where FWI would
normally fail, since the result improvement easily offsets the extra
computations. Naturally, hybrid combinations of WRI$\ast$ and FWI are
also possible, as shown here, to amortize these costs.

\section{Conclusion}\label{conclusion}

We proposed a seismic inversion method whose main strength is being both
robust with respect to local minima and scalable to 3D. Traditional
methods are either scalable but prone to cycle skipping, like full
waveform inversion, or less sensitive to the starting model but
unfeasible for large problems, like wavefield reconstruction inversion.
Our proposal, despite being a reformulation of wavefield reconstruction
inversion, can leverage on modern time-domain solvers, which is key to
tackle realistically sized problems. Numerical experiments indicate that
our proposal is consistently superior to full waveform inversion when
cycle skipping occurs. Moreover, it benefits from a relaxed formulation
of the wave equation, and is more resilient than full waveform inversion
with respect to starting models when inaccurate modeling assumptions are
made. A comparison with conventional wavefield reconstruction inversion
highlights competitive results, albeit with some loss of resolution,
depending on the choice of hyperparameters like the estimated data noise
level. It has been shown that these shortcomings can be easily avoided
with an hybrid scheme involving full waveform inversion and/or
pseudo-Hessian preconditioning. Our method is mature for 3D, but the
computational effort required is twice as much what is prescribed by
full waveform inversion (per gradient calculation). This extra cost can
be reduced by aforementioned hybrid schemes and stochastic optimization.

\section{Related material}\label{related-material}

A Julia implementation of the method herein described can be found in
the following repository:

\url{https://github.com/slimgroup/JUDI.jl/tree/master/examples/twri}

\section{Acknowledgments}\label{acknowledgments}

This work has been implemented in Julia and leverages the Devito
framework for seismic modeling, which allows automatic generation of
highly-optimized finite-difference C code, given only a symbolic
representation of the wave equation
\citep{louboutin2018devito, luporini2018architecture}. From Julia, we
interface to Devito through the JUDI package \citep{witte2019large}.

\bibliography{wrid_arxiv}

\end{document}